\documentclass[traditabstract]{aa} 

\usepackage{graphicx}
\usepackage{bm}
\usepackage{natbib}
\usepackage{txfonts}

\def\chandra    {{\em Chandra}\/}

\begin{document}
   \title{Search for cold and hot gas in the ram pressure stripped 
Virgo dwarf galaxy IC3418\thanks{Based on observations carried out with 
the IRAM 30m Telescope and with the Chandra X-ray Observatory. IRAM is 
supported by INSU/CNRS (France), MPG (Germany), and IGN (Spain)}}
   \author{P. J\'achym\inst{1,2},
          J. D. P. Kenney\inst{2},
          A. R\accent23 u\v zi\v cka\inst{1},
          M. Sun\inst{3},
          F. Combes\inst{4},
          \and 
          J. Palou\v s\inst{1}
          }

   \institute{Astronomical Institute, Academy of Sciences of the Czech Republic,
              Bo\v cn\'i II 1401, 141 00 Prague, Czech Republic\\
              \email{jachym@ig.cas.cz}
         \and
             Department of Astronomy, Yale University, 260 Whitney Ave., New Haven, CT 06511, U.S.A.\\
             \email{jeff.kenney@yale.edu}
         \and
             Eureka Scientific, 2452 Delmer Street Suite 100, Oakland, CA 94602, U.S.A.\\
             \email{mingsun.cluster@gmail.com}
         \and
             Observatoire de Paris, LERMA, 61 Av. de l'Observatoire, 75014 Paris, France\\
             \email{francoise.combes@obspm.fr}
             }

   \date{Received October 4, 2012; accepted April 30, 2013}

\abstract{
We present IRAM 30m sensitive upper limits on CO emission in the ram 
pressure stripped dwarf Virgo galaxy IC3418 and in a few positions 
covering \ion{H}{ii} regions in its prominent 17~kpc UV/H$\alpha$ 
gas-stripped tail. In the central few arcseconds of the galaxy, we 
report a possible marginal detection of about $1\times 10^6~M_\odot$ of 
molecular gas (assuming a Galactic CO-to-H$_2$ conversion factor) that 
could correspond to a surviving nuclear gas reservoir. We estimate that 
there is less molecular gas in the main body of IC3418, by at least a 
factor of 20, than would be expected from the pre-quenching UV-based 
star formation rate assuming the typical gas depletion timescale of 
2~Gyr. Given the lack of star formation in the main body, we think the 
H$_2$-deficiency is real, although some of it may also arise from a 
higher CO-to-H$_2$ factor typical in low-metallicity, low-mass 
galaxies. The presence of \ion{H}{ii} regions in the tail of IC3418 
suggests that there must be some dense gas; however, only upper limits 
of $< 1\times 10^6~M_\odot$ were found in the three observed points in 
the outer tail. This yields an upper limit on the molecular gas content 
of the whole tail $< 1\times 10^7~M_\odot$, which is an amount similar 
to the estimates from the observed star formation rate over the tail. 
We also present strong upper limits on the X-ray emission of the 
stripped gas in IC3418 from a new \chandra\ observation. The measured 
X-ray luminosity of the IC3418 tail is about 280 times lower than that 
of ESO~137-001, a spiral galaxy in a more distant cluster with a 
prominent ram pressure stripped tail. Non-detection of any diffuse 
X-ray emission in the IC3418 tail may be due to a low gas content in 
the tail associated with its advanced evolutionary state and/or due to 
a rather low thermal pressure of the surrounding intra-cluster medium. 
}

   \authorrunning{P. J\'achym et al.}
   \titlerunning{Search for cold and hot gas in IC3418}
   \maketitle

%

\section{Introduction}
The evolution of galaxies in clusters is driven by interactions between 
the interstellar medium (ISM) of the galaxies and the gas in the 
intra-cluster medium (ICM) \citep{gunngott1972, poggianti1999, 
vangorkom2004, koopmann2004}. Because of its proximity and dynamical 
youth \citep{schindler1999, randall2008}, the Virgo Cluster is an 
excellent laboratory for studying the processes that transform galaxies 
in dense environments. By far the most numerous galaxies in Virgo are 
the dwarfs. The relationship between the blue star-forming dwarf 
irregulars (dIs) and the more common dwarf ellipticals (dEs) has been a 
long-standing puzzle. Dwarf ellipticals are centrally concentrated in 
the cluster, whereas blue star-forming dwarf irregulars are 
preferentially located in the outskirts, although their masses and 
structural parameters are similar \citep{binggeli1987}. It has 
therefore been suggested that dEs might transform from gas-rich dwarf 
galaxies which lose their interstellar gas and subsequently quench 
their star formation because of external perturbations from the 
surrounding environment \citep{boselli2008}, such as ram pressure 
stripping. Several authors have discussed this and other evidence for 
dwarf galaxy transformation through ram pressure stripping, but until 
recently there has not been any clear case of a galaxy actually 
undergoing such a transformation. We think IC3418 is such a case. 

\begin{figure}[t]
\centering
\includegraphics[width=0.45\textwidth]{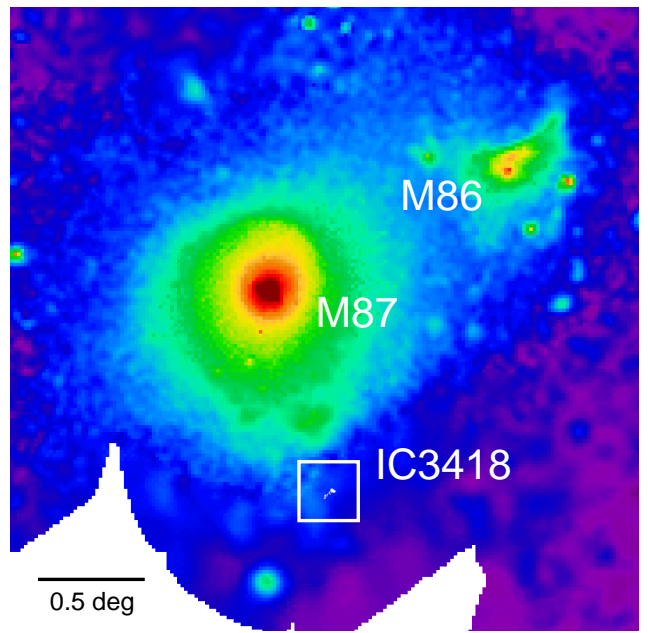}
\caption{
The location of IC3418 in the Virgo cluster. The background image shows the 
distribution of the $0.5-2.0$~keV ICM as observed in X-rays with ROSAT 
(Credit: S. L. Snowden, http://heasarc.gsfc.nasa.gov). Silhouette of a 
GALEX UV image of IC3418 is displayed at the position of the galaxy.
}\label{FigVirgoA}
\end{figure}

The galaxy IC3418 (VCC1217) is a peculiar dwarf galaxy located very 
close to the cluster core, with a projected distance of only $1\degr$ 
($\sim 290$~kpc) from M87 (see Fig.~\ref{FigVirgoA}), and a 
line-of-sight velocity with respect to the cluster mean of about 
-1000~km\,s$^{-1}$. It is very likely on its first orbit through the 
Virgo center, and probably close to its nearest approach to M87. No 
\ion{H}{i} was detected anywhere in the galaxy down to a $3\sigma$ 
limit of $\sim 8\times 10^6~M_\odot$ per beam \citep{chung2009}; in the 
main body of the galaxy there is also no H$\alpha$ emission 
\citep{hester2010}. The stellar metallicity of the galaxy was measured 
from optical spectroscopy to be $\sim 0.5\pm 0.2~Z_\odot$ (Kenney et 
al., in prep.). A new stellar population analysis indicates that star 
formation in the galaxy stopped about $250\pm 50$~Myr ago in the 
central $25\arcsec$, and that it took $\la 70$~Myr for star formation to 
cease from $\sim 40\arcsec$ to the center (Kenney et al., in prep.). 
This is consistent with ram pressure suppression of star formation from 
the outside inwards, acting typically on a timescale of $\sim 100$~Myr. 
The main stellar body of the galaxy is symmetric and undisturbed 
indicating that a tidal interaction is excluded (Kenney et al., in 
prep.). 

\begin{table}
\caption[]{Parameters of IC3418 (VCC~1217).} \label{ic3418}
\begin{tabular}{ll}
\hline
\hline
\noalign{\smallskip}
RA, Dec (J2000) 		& 12:29:43.8, +11:24:09\\
type 				& IBm\\
$\upsilon_{\rm helio}$\tablefootmark{a} 	& $170\pm 10$~km~s$^{-1}$\\
$D_{\rm M87}$ projected		& $1.02\degr$ ($\approx290$~kpc)\tablefootmark{h}\\
major/minor diameter\tablefootmark{b}	& $1.5\arcmin/1\arcmin$ 
($\approx7.2$~kpc/4.8~kpc)\tablefootmark{h}\\
PA, inclination\tablefootmark{c}	& $\sim 45\degr$, $\sim 50\degr$\\
apparent/absolute $B$ mag.\tablefootmark{d}	& 14.85 / -16.2\\
apparent/absolute $K$ mag.\tablefootmark{e}	& 12.55 / -18.5\\
FUV, NUV, FUV-NUV\tablefootmark{f}	& 17.55, 16.69, 0.85$\pm 0.10$\\
\ion{H}{i} mass\tablefootmark{g} 		& $<8\times 10^6~M_\odot$ ($3\sigma$)\\
\ion{H}{i}-deficiency\tablefootmark{g} 	& $>2.16$\\
tail projected length		& $3.5\arcmin$ ($\approx17$~kpc)\tablefootmark{h}\\
\hline
\noalign{\smallskip}
\end{tabular}
\tablefoot{
\tablefoottext{a}{Kenney et al. (in prep.); 
38~km\,s$^{-1}$ in GoldMine \citep{gavazzi2003};} 
\tablefoottext{b}{NED}; 
\tablefoottext{c}{HyperLEDA \citep{paturel2003};} 
\tablefoottext{d}{Internal and Galactic extinction corrected 
\citep{gavazzi2006}. Corresponding $B$-band luminosity is $4.8\times 
10^8~L_\odot$, using $M_{B,\odot}= 5.48$ \citep[e.g.,][]{mo2010};}
\tablefoottext{e}{GoldMine. Foreground Galactic extinction in the 
$K$-band is 0.012~mag (NED). Corresponding $K$-band luminosity is 
$5.2\times 10^8~L_\odot$, using $M_{K,\odot}=3.28$ 
\citep[e.g.,][]{mo2010};}
\tablefoottext{f}{\citet{gildepaz2007};} 
\tablefoottext{g}{\citet{chung2009};} 
\tablefoottext{h}{Assuming Virgo cluster distance of 16.5 Mpc 
\citep{mei2007}.}}
\end{table}

A ram pressure stripping scenario for IC3418 is also suggested by 
optical and GALEX UV observations \citep{chung2009, hester2010, 
fumagalli2011} that revealed a remarkable one-sided tail extending 
17~kpc from the galaxy's SE side. It comprises about nine bright knots 
and linear, parallel filaments of young UV-bright stars. Only in the 
outer half of the tail H$\alpha+[$\ion{N}{ii}] emission is found, 
mostly associated with fireballs \citep{yoshida2008}, elongated 
streams of young stars with \ion{H}{ii} regions and bright UV knots at 
their heads (Kenney et al., in prep.). These are interpreted as dense, 
star-forming gas clouds which are accelerated by ram pressure, leaving 
behind trails of newly-formed stars that are not affected by ram 
pressure and therefore decouple from the gas. 

While good examples of ram pressure stripping are known among the Virgo 
spiral galaxies \citep[e.g.,][]{kenney2004, chung2007, chung2009, 
abramson2011}, none of these galaxies is completely gas stripped and 
none has such a luminous UV tail, or linear parallel stellar streams, 
or fireballs as IC3418. There are already known examples of one-sided 
tails of young stars extending from more massive galaxies in more 
distant, richer clusters \citep{cortese2007, yoshida2008, smith2010, 
sun2010, woudt2008}, where the ram pressure can be 1 to 2 orders of 
magnitude stronger than in Virgo, but IC3418 is by far the closest such 
galaxy known. 

\begin{figure}[t]
\centering
\includegraphics[width=0.45\textwidth]{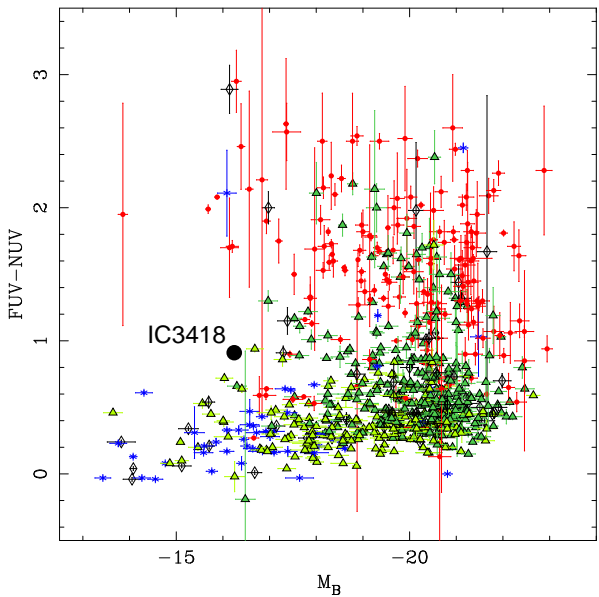}
\caption{
The location of IC3418 in the $({\rm FUV-NUV}) - M_B$ color-magnitude 
diagram of nearby galaxies from the sample of \citet{gildepaz2007}. 
${\rm FUV-NUV}= 0.85\pm0.10$ \citep{gildepaz2007} and $m_B$ corrected 
for internal and Galactic extinction is 14.85~mag (NED), i.e., $M_B= 
-16.2$. Red dots are elliptical/lenticular galaxies, dark green 
triangles are early-type spirals, light green triangles are late-type 
spirals, blue asterisks are irregular and compact galaxies, and black 
diamonds are galaxies lacking morphological classification. Figure 
adopted from \citet{gildepaz2007}.
}\label{FigGildePaz}
\end{figure}

In Fig.~\ref{FigGildePaz}, IC3418 is placed into a $({\rm FUV-NUV})$ 
vs. $M_B$ color-magnitude diagram of a sample of nearby galaxies 
studied by \citet{gildepaz2007}. It shows that IC3418 stands outside 
both the red and blue sequences of galaxies, respectively, and 
that it occurs right at the boundary between early- and late-type 
galaxies. This suggests that IC3418 is being transformed from a dwarf 
irregular type to an early-type galaxy (see Kenney et al., in prep.) 
which is consistent with the recent cease of its star formation due to 
ram pressure stripping. 

Numerical modeling has shown that ram pressure stripping is a very 
efficient process for removing the gaseous content of cluster galaxies 
\citep[e.g.,][]{vollmer2001, roediger2005, jachym2007, jachym2009}. 
Simulations also predict that material in the gas-stripped tails gets 
compressed by ram pressure and radiative cooling can form new molecular 
clouds in situ, which might then form a population of new stars 
\citep{kapferer2009, tonnesen2009}. Dwarf galaxies have weaker 
potential wells and their ambient ISM may be correspondingly less dense 
\citep{bolatto2008}. This, plus a strong ram pressure near the cluster 
center, could significantly enhance stripping of IC3418 and thus 
account for its peculiar morphology. The presence of star formation in 
the tail, strongly suggesting the presence of molecular gas, makes 
IC3418 an attractive place to search for CO emission in the 
environmentally affected galaxy. However, dwarf irregulars are usually 
fainter in CO than large spirals when normalized by (stellar) mass and 
those with metallicities less than $\sim 8.0$ (in oxygen abundance 
scale) are challenging to detect \citep[e.g.,][]{taylor1998, 
komugi2011}. 

In this paper we want to address two main questions: (1) Is the main 
body of IC3418 totally stripped or did some gas survive the stripping? 
We present the results of our IRAM 30m search for CO emission since 
molecular gas is more likely to survive than \ion{H}{i}. (2) How does 
the ISM behave in ram pressure stripped, star-forming tails? We are 
interested in searching for molecular and for hot X-ray emitting gas in 
the tail in order to help characterize the ISM in such tails. We 
present the results of our new sensitive \chandra\ observation of the 
galaxy. 

The structure of the paper is as follows: after a brief description of 
our observations (Sect.~\ref{observations}), we present and analyze our 
IRAM 30m results (Sect.~\ref{resultsiram}), and estimate upper limits 
on the current molecular gas content of the whole IC3418 
(Sect.~\ref{current}). We then estimate the original molecular gas 
content of the galaxy from the star formation rate in the main body and 
compare the molecular content of IC3418 with other low-mass galaxies 
(Sect.~\ref{original}). The H$_2$-deficiency of the galaxy is also 
discussed. In Sect.~\ref{resultschandra} we analyze the \chandra\ 
results. Then, by means of numerical calculations we study the effects 
of ram pressure on ISM with different column density (Sect.~\ref{RPS}). 
In Sect.~\ref{discussion} we discuss the structure of the gas-stripped 
tail, as well as the fate of the stripped gas, and origin of star 
formation in the tail. The conclusion follows in 
Sect.~\ref{conclussion}. 

\section{Observations}\label{observations}
\subsection{Millimeter integration: IRAM 30m}
\begin{figure}[t]
\centering
\includegraphics[height=0.45\textwidth]{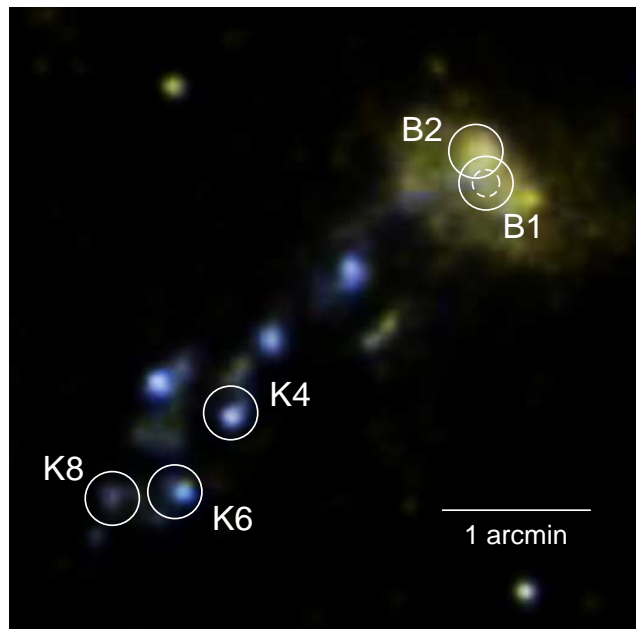}
\caption{
Observed positions on a GALEX image (FUV in blue, NUV in yellow) 
of IC3418 with $^{12}$CO(1-0) beams (${\rm FWHM}= 21\arcsec$) of the 30m 
IRAM telescope displayed -- two positions (B1, B2) in the main body of 
the galaxy and three (K4, K6, K8) in the outer part of the tail. The 
CO(2-1) beams have the same centers but half the diameter as the 
circles shown -- a CO(2-1) beam is shown in B1 with a dashed circle. 
The background image is adopted from \citet{hester2010}. 
}\label{Figiram}
\end{figure}

Using the 30m telescope operated by the Institut de Radio Astronomie 
Millim\'etrique (IRAM) at Pico Veleta, Spain, in August 2011 we carried 
out observations towards the main body of IC3418 and the brightest 
H$\alpha$/UV regions of its tail (see scheme in Fig.~\ref{Figiram}). We 
used the EMIR receiver in E090 and E230 bands to observe simultaneously 
at the frequencies of the $^{12}$CO(1--0) ($\nu_{\rm rest} = 
115.271$~GHz) and the $^{12}$CO(2--1) ($\nu_{\rm rest} = 230.538$~GHz) 
lines. At these frequencies the telescope half-power beamwidths are 
$21\arcsec$ and $11\arcsec$, respectively, which corresponds to a resolution of 
about 1.7~kpc and 0.9~kpc, respectively, at the adopted distance of the 
Virgo cluster of 16.5~Mpc \citep{mei2007}. As backends, we used the 
4~MHz filterbanks at 115~GHz with 10.4~km\,s$^{-1}$ velocity 
resolution, and the WILMA autocorrelator with a spectral resolution of 
2~MHz at both 115~GHz and 230~GHz (i.e., 5.2~km\,s$^{-1}$ velocity 
resolution at 115~GHz and 2.6~km\,s$^{-1}$ at 230~GHz). The new FTS 
spectrometer with 0.192~kHz spectral resolution was also connected to 
both lines as a back-up. In this work, WILMA and 4MHz data are 
analyzed. 

All observations were done in the Wobbler switching mode that is known 
to provide much flatter baselines, with ripples at a much lower 
amplitude than with the position switching mode. The secondary 
reflector was switching symmetrically at 0.7~kHz frequency with 
$120\arcsec$ 
amplitude in azimuth. The actual distance between the source and a 
reference position is double the amplitude value, i.e., $240\arcsec$, which 
is large enough to just avoid the tail with the standard azimuthal 
direction of switching. 

The main beam efficiencies $B_{\rm eff}$ of the 30m antenna are 
estimated to 0.79 at 115~GHz and 0.57 at 230~GHz, and the forward 
efficiencies $F_{\rm eff}$ to 0.94 and 0.92, respectively. The 
corrected antenna temperatures provided by the telescope were converted 
to the main beam brightness temperature by $T_{\rm mb} = T_{\rm A}^*\, 
F_{\rm eff}/ B_{\rm eff}$. To check the system setup we observed 
IRC+10216 and M99 and we obtained lines matching the catalogued lines. 

Weather conditions were typical for the summer afternoon period at Pico 
Veleta with the PWV exceeding 10~mm in the first half of the run. Then, 
after a passage of a weather front, the conditions improved 
substantially and PWV decreased to $5-7$~mm. The atmosphere was, 
however, rather unstable and pointing and focusing were difficult. 
Pointing was checked about every 1.5~hours mostly on Saturn which was 
close to our sources, usually with small corrections from $2\arcsec$ to 
$5\arcsec$. The system temperatures are given in Table~\ref{Sources}. 

\begin{table}
\caption[]
{List of observed positions. System temperatures are given in $T_{\rm 
A}^*$ scale.
}
\label{Sources}
\begin{tabular}{lllllll}
\hline
\hline
\noalign{\smallskip}
 & R.A.    & Dec.    & $T_{\rm ON}$ & $T_{\rm sys}^{(1-0)}$ & $T_{\rm sys}^{(2-1)}$\\
 & (J2000) & (J2000) & (hr)         & (K)                   & (K)      \\
\noalign{\smallskip}
\hline
\noalign{\smallskip}
B1 & 12:29:43.8  & +11:24:09    & 5.6 & 240-300 & 400-500\\
B2 & 12:29:44.07 & +11:24:22.38 & 2.8 & 200-300 & 230-400\\
K4 & 12:29:50.75 & +11:22:36.35 & 2.5 & 250-350 & 400-700\\
K6 & 12:29:52.29 & +11:22:04.86 & 2.7 & 250-330 & 500-600\\
K8 & 12:29:53.92 & +11:22:01.85 & 1.7 & 210-280 & 240-310\\
\noalign{\smallskip}
\hline
\end{tabular}
\end{table}

The sky coordinates of the observed positions shown in Fig.~\ref{Figiram} 
are given in Table~\ref{Sources}, together with respective on-source 
times. These were typically 2.5~hr per position per polarization, 
divided into 6~min scans. For the B1 position it was roughly double. 
For K8, the on-source time was shorter by a factor of $\sim 1.5$ than 
for the K4 and K6, but observing conditions had improved resulting in 
a better sensitivity. The data were reduced with CLASS (Continuum and 
Line Analysis Single-dish Software) developed by IRAM in the standard 
manner: the spectra were checked for errors (spikes) and bad channels 
exceeding the $5\sigma$ level were flagged. The quality of WILMA and 
4~MHz data was very good. 
Linear baselines were fitted. Each spectrum was re-binned to a velocity 
resolution of 10.4~km\,s$^{-1}$ and both polarizations were averaged in 
order to increase the signal-to-noise (S/N) ratio. All scans were then 
summed up to produce the final spectra. The root mean square (rms) 
noise levels of about $1-2$~mK at CO(1-0) frequency were achieved in 
the final summed spectra (Table~\ref{Results}). 

\subsection{X-ray imaging: Chandra}
The observation of IC3418 was performed with the Advanced CCD Imaging 
Spectrometer (ACIS) on November 12, 2012 (obsID: 13811). Standard 
\chandra\ data analysis was performed which includes the corrections 
for the slow gain change, charge transfer inefficiency, and the ACIS 
low-energy quantum efficiency degradation from the contamination on 
the ACIS optical blocking filter. No flares of particle background were 
present in the observation. The effective exposure time is 33.8~ks for 
the S3 chip where IC3418 is positioned. CIAO4.4 was used for the data 
analysis. The calibration files used correspond to \chandra\ 
calibration database 4.5.3 from the \chandra\ X-ray Center. The solar 
photospheric abundance table by \citet{anders1989} were used in 
the spectral fits. We adopted an absorption column density of 
$2.2\times 10^{20}$~cm$^{-2}$ from the Leiden/Argentine/Bonn \ion{H}{i} survey 
\citep{kalberla2005}. 

\section{Results -- IRAM}\label{resultsiram}
In a set of deep integrations we observed two positions in the main 
body of the galaxy where dense molecular gas might have survived the 
effects of the cluster environment, and three positions in the outer 
tail where the presence of dense gas is suggested from the existence of 
\ion{H}{ii} regions (Fig.~\ref{Figiram}, see coordinates in 
Table~\ref{Sources}). The B1 pointing is centered at the galaxy's 
optical center (as given by NED), and the B2 point is shifted by less 
than one CO(1-0) beamwidth in the NE direction, towards the brightest 
region of the galaxy where significant substructure occurs, indicating 
that star formation occurred there recently. The tail positions K4, K6, 
and K8\footnote{
The naming of the tail positions corresponds to \citet{hester2010}. 
\citet{fumagalli2011} use a different notation (K-6, K-4, K-3).} are 
associated with peaks in UV and H$\alpha$ emission.

\subsection{Main body}\label{irammain}
\begin{figure*}
\centering
\includegraphics[height=0.33\textwidth,angle=270]{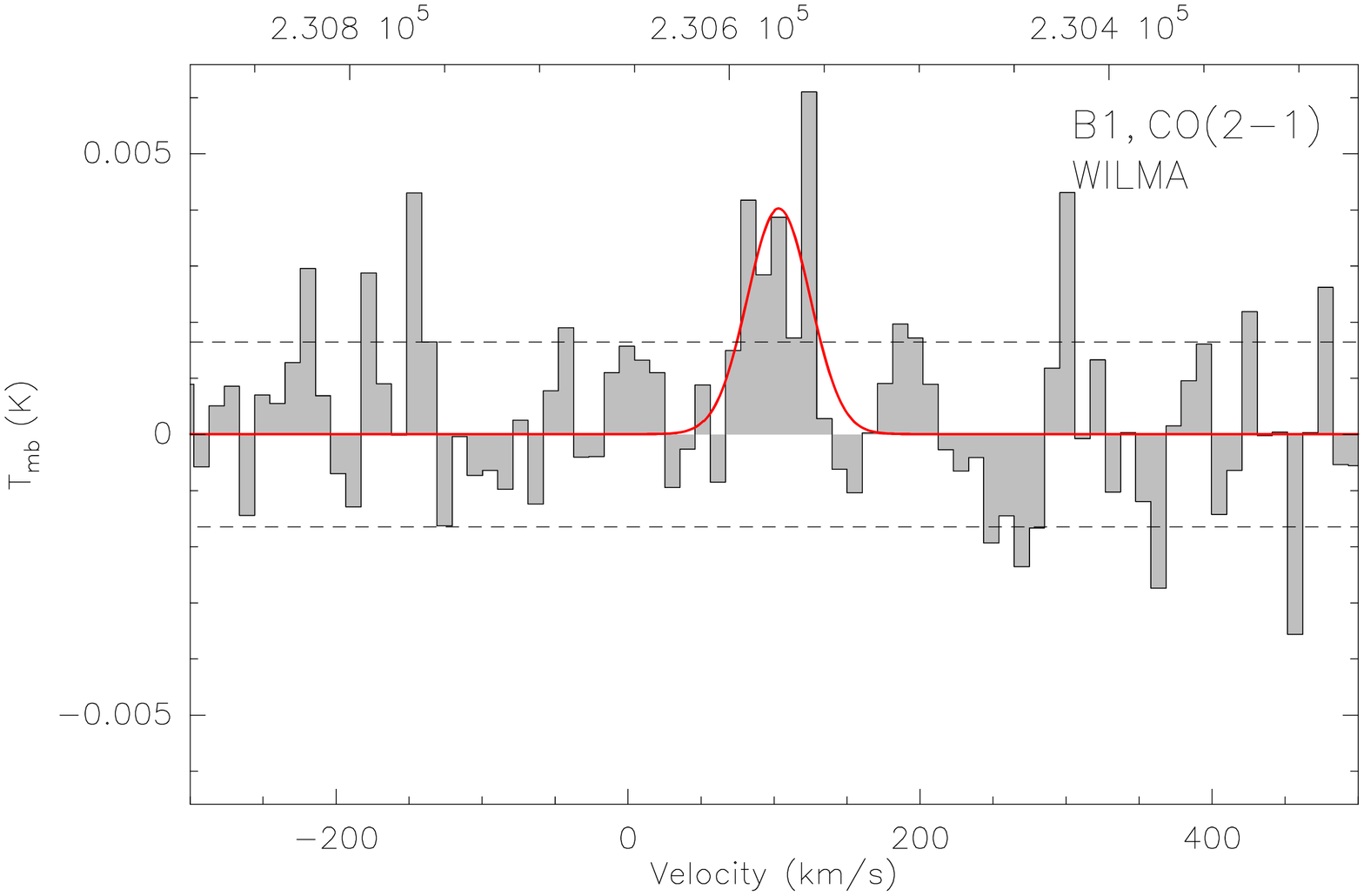}
\includegraphics[height=0.33\textwidth,angle=270]{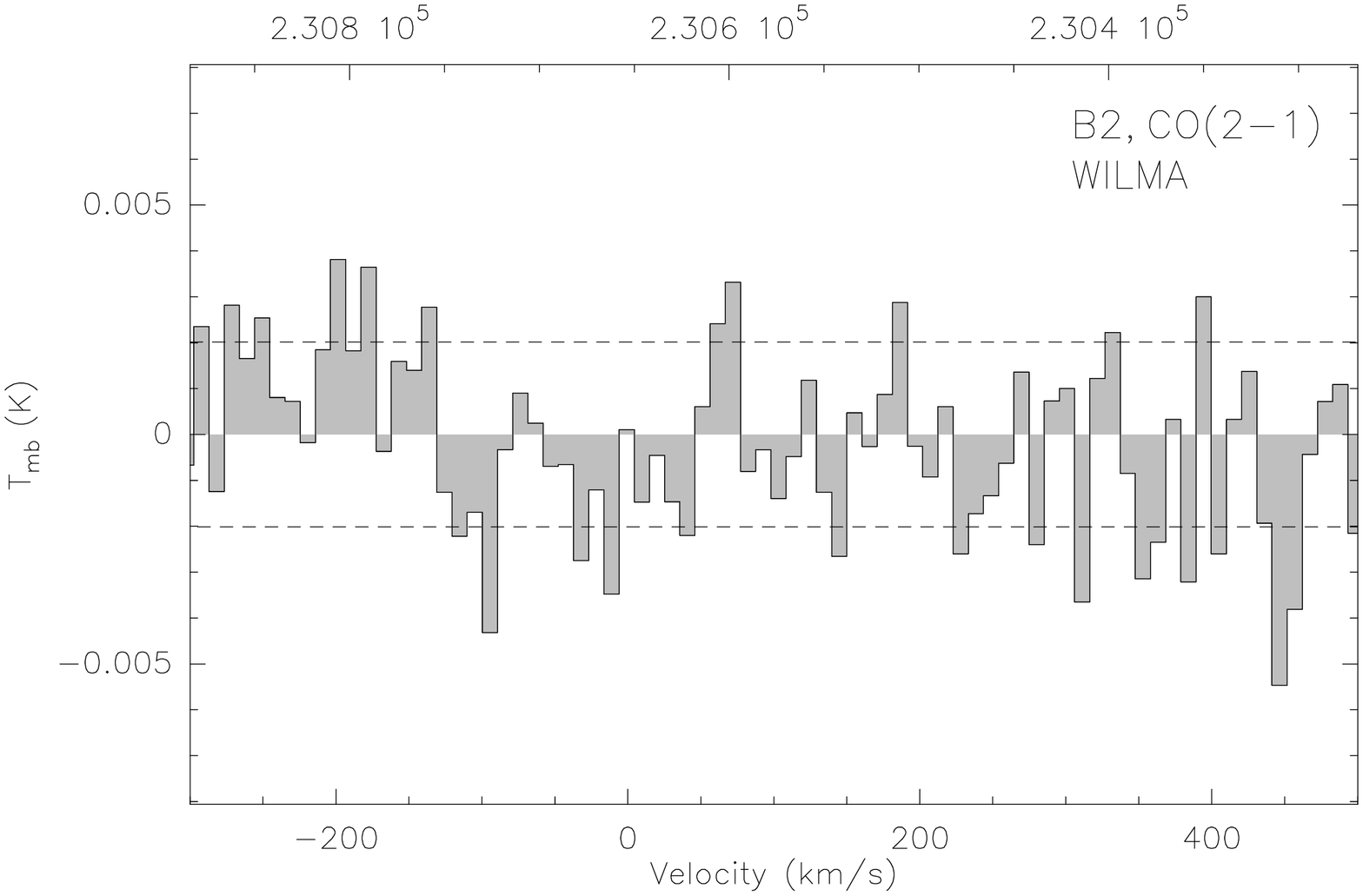}
\includegraphics[height=0.33\textwidth,angle=270]{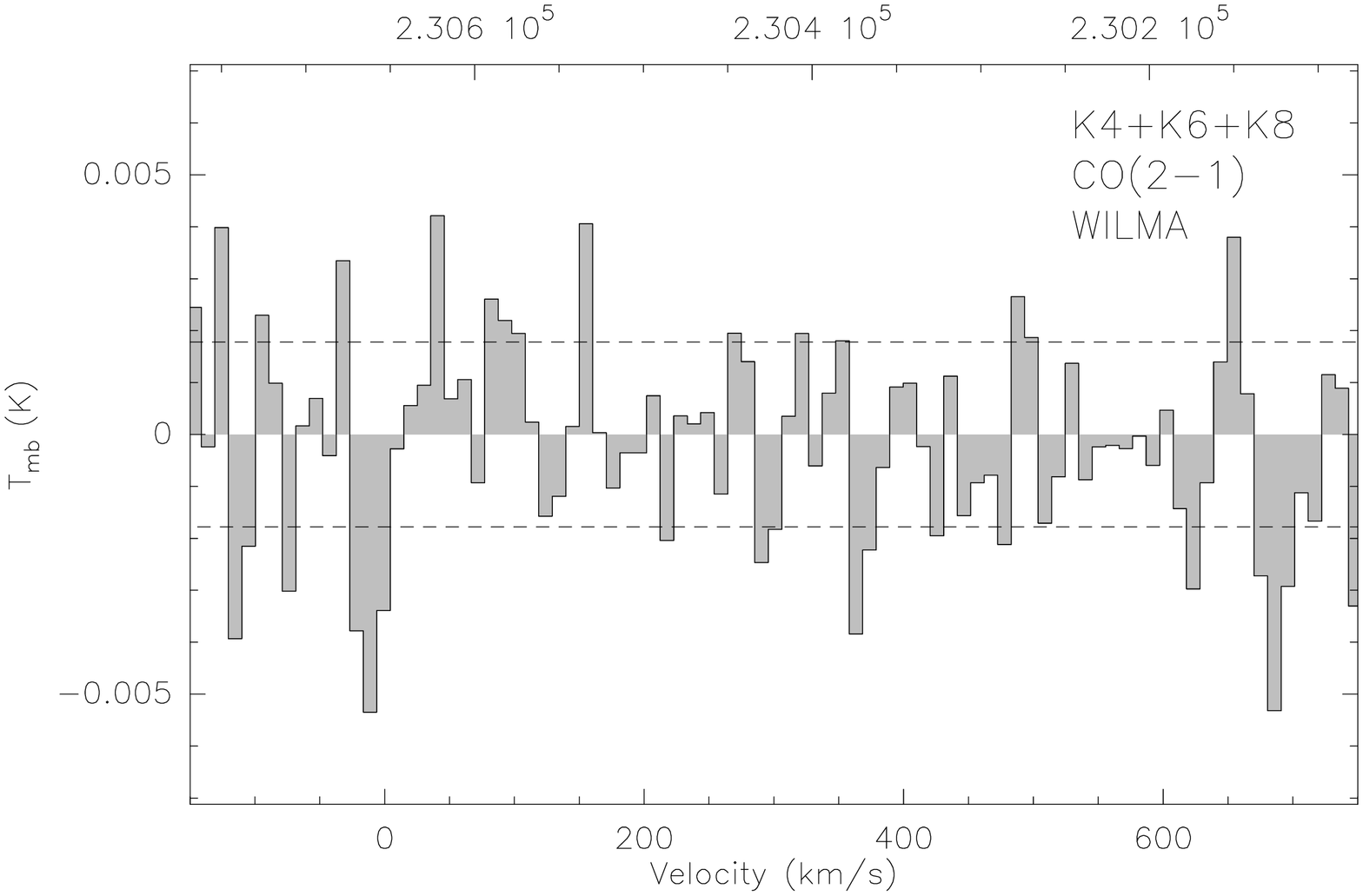}
\includegraphics[height=0.33\textwidth,angle=270]{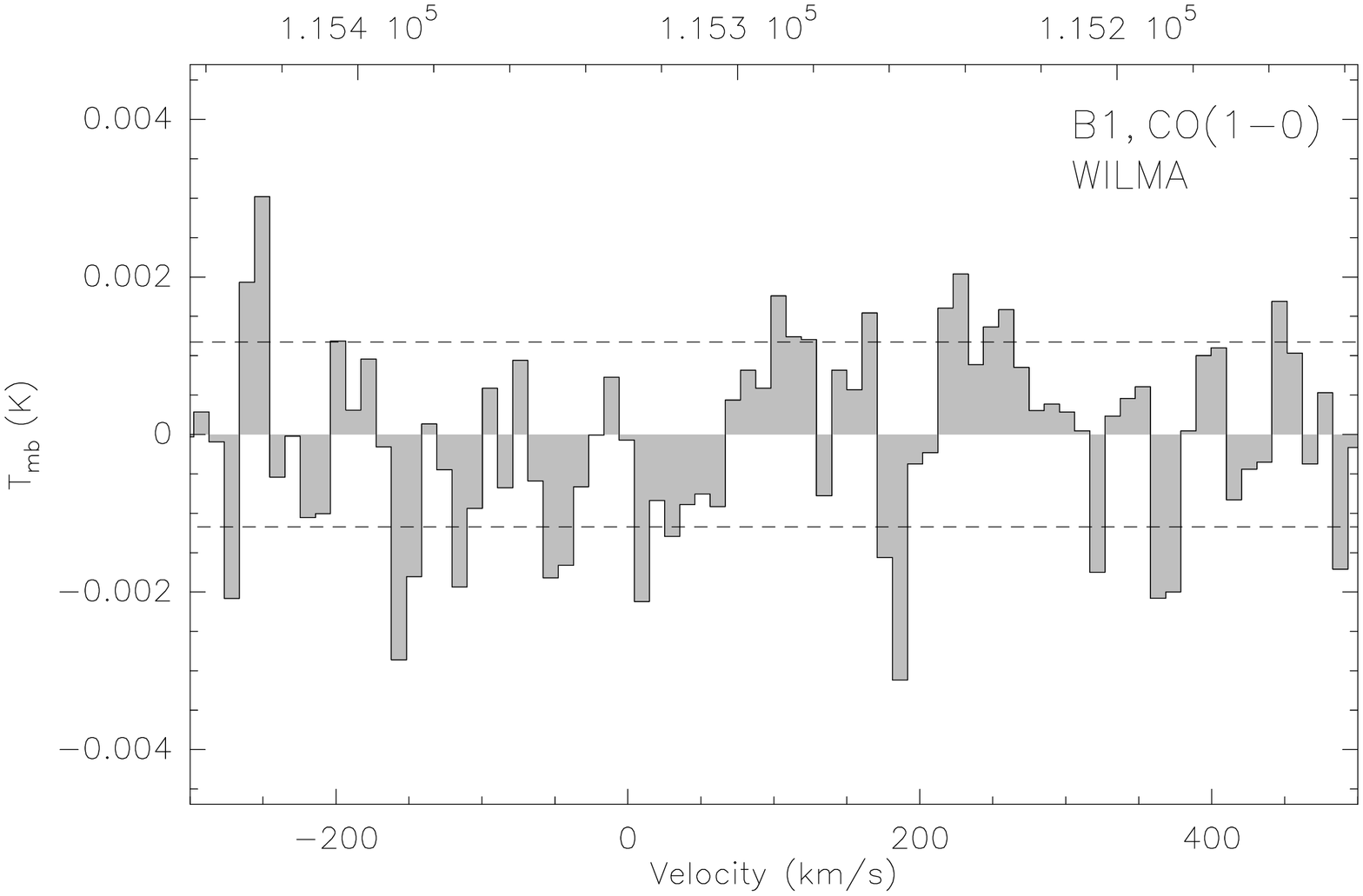}
\includegraphics[height=0.33\textwidth,angle=270]{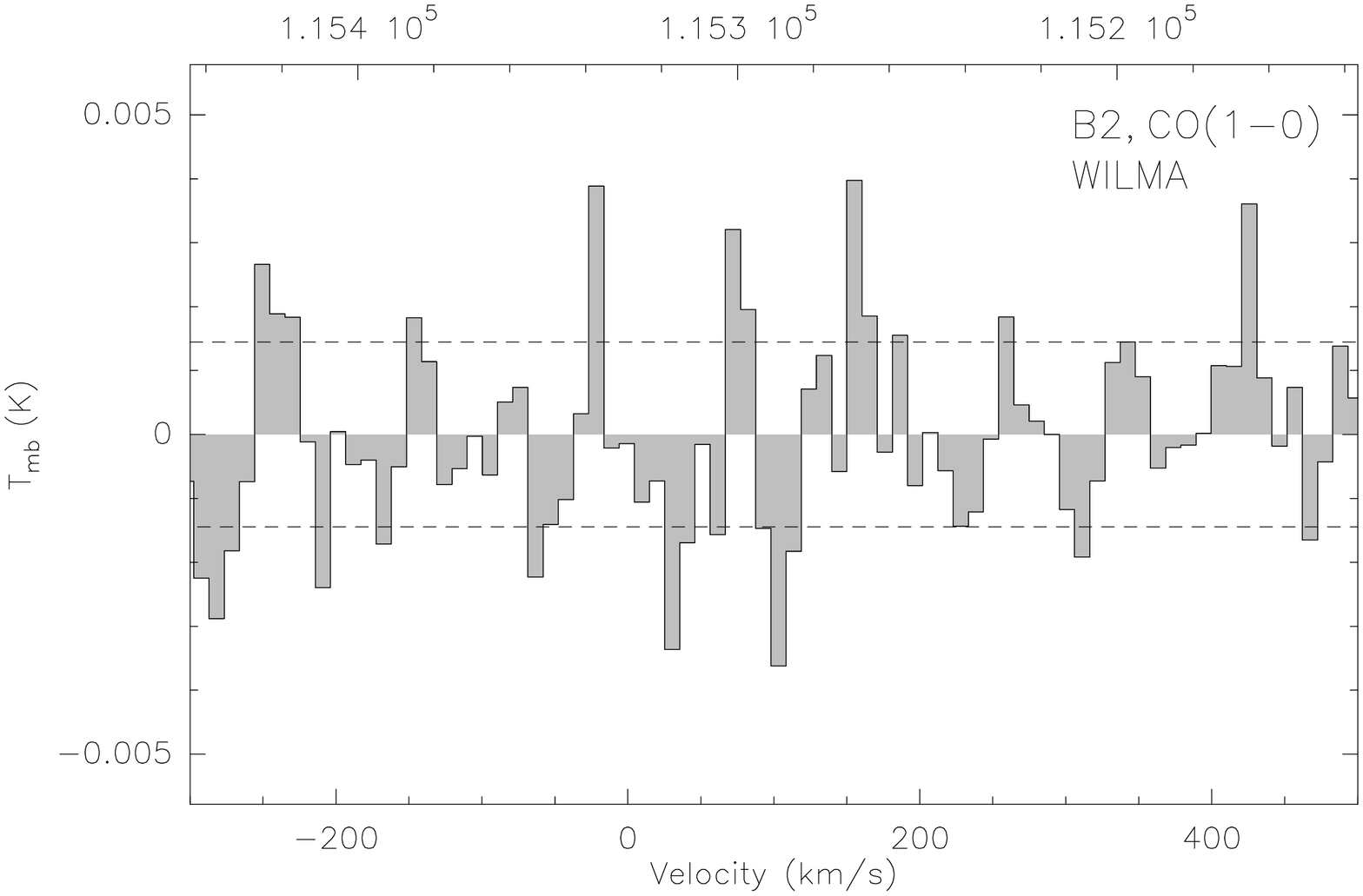}
\includegraphics[height=0.33\textwidth,angle=270]{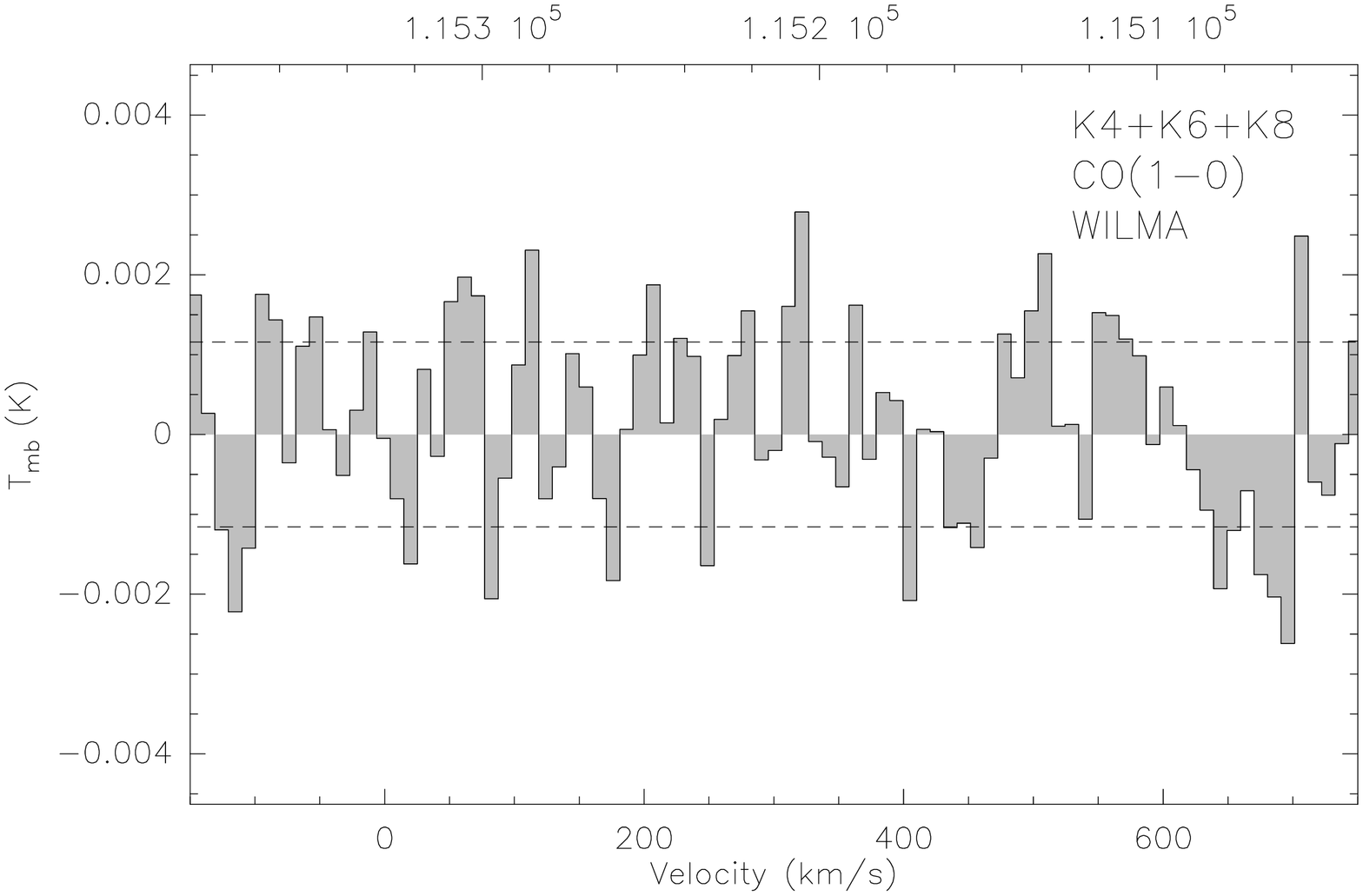}
\caption{
Marginal detection in the central main body position B1 ({\it left}), 
spectra in the off-center position B2 ({\it middle}), and combined 
K4+K6+K8 tail spectra ({\it right}): WILMA CO(2-1) ({\it top}) and 
WILMA CO(1-0) ({\it bottom}) spectra smoothed to 10.4~km\,s$^{-1}$ 
resolution. Dashed lines show $\pm 1\sigma$ noise levels; $y$-axis 
scaling is set to $\pm 4\sigma$. The velocity scale is 
LSR; for the sky position of IC3418, $\upsilon_{\rm helio}= 
\upsilon_{\rm LSR}- 3.2$~km\,s$^{-1}$. 
}\label{FigB1B2K}
\end{figure*}

First we searched for CO emission in the central position B1. This was 
our deepest integration with the total on-source time of about 5.6~hr 
(per polarization). The left panels in Fig.~\ref{FigB1B2K} show WILMA 
CO(2-1) and WILMA CO(1-0) spectra smoothed to 10.4~km\,s$^{-1}$ 
resolution. Very low rms values of about 1.7~mK and 1.1~mK, 
respectively, were achieved. Although no strong detection appears, a 
feature that is six channels wide ($\sim 60$~km\,s$^{-1}$) occurs at 
both frequencies at the central velocity of about 100~km\,s$^{-1}$. It 
is more prominent in the CO(2-1) spectrum. We inspected the surrounding 
parts of the spectra and found no other similar case. The 
signal-to-noise ratios of the line feature in the WILMA CO(1-0) and 
WILMA CO(2-1) spectra are 2.1 and 5.6, respectively, where we have 
calculated ${\rm S/N}= N^{-1/2}\ \sigma_{\rm rms}^{-1}\ \sum_{i=1}^N 
I_i$, where $N$ is the number of channels covered by the line, 
$\sigma_{\rm rms}$ is the rms intensity of the spectrum per channel, 
and $I_i$ is the brightness temperature in the $i$-th channel. In order 
to claim a detection, we require the integrated intensity $I_{\rm CO}= 
\int T_{\rm mb}\, d\upsilon$ to be at least three times greater than 
the noise over the spectral channels covered by the line\footnote{The 
noise over $N$ spectral channels covered by the line $\sigma_{\rm I}= 
\Delta \upsilon_{\rm res}\ \sigma_{\rm rms} N^{1/2}= \sigma_{\rm rms} 
(\Delta \upsilon_{\rm res} \langle \Delta \upsilon_{\rm CO} 
\rangle)^{1/2}$, where $\Delta \upsilon_{\rm res}$ is the velocity 
resolution and $\langle \Delta \upsilon_{\rm CO} \rangle$ is the mean 
FWHM linewidth.}. 

\begin{table*}[t]
\centering
\caption[]
{Properties of the observed positions in IC3418.}
\label{Results}
\begin{tabular}{ccccccccc}
\hline\hline
\noalign{\smallskip}
Source & rms$_{\rm W(1-0)/ W(2-1)/ 4MHz}$ & FWHM & $I_{\rm CO(1-0)}$ & 
$I_{\rm CO(2-1)}$ & $L_{\rm CO(1-0)}$ & $L_{\rm CO(2-1)}$ & $M_{\rm 
mol}^{\rm CO(1-0)}$ & $M_{\rm mol}^{\rm CO(2-1)}$\\
       & (mK) & (km\,s$^{-1}$) & (K\,km\,s$^{-1}$) & (K\,km\,s$^{-1}$) 
& ($10^5$~K\,km\,s$^{-1}$\,pc$^2$) & ($10^5$~K\,km\,s$^{-1}$\,pc$^2$) & 
($10^6~M_\odot$)& ($10^6~M_\odot$)\\
(1) & (2) & (3) & (4) & (5) & (6) & (7) & (8) & (9)\\
\noalign{\smallskip}
\hline
\noalign{\smallskip}
B1 & 1.1 / 1.6 / 1.0 & $54\pm 12$ &  $0.06\pm 0.03$ & $0.24\pm 0.05$ & $2.1$ & $2.1$ & $0.9$ & $1.2$\\
B2 & 1.4 / 2.0 / 1.3 & $\sim 30$  & $<0.1$ & $<0.1$ & $<2.7$ & $<0.9$ & $<1.2$ & $<0.5$\\
K4 & 2.0 / 3.6 / 1.8 & $\sim 30$  & $<0.1$ & $<0.2$ & $<3.8$ & $<1.7$ & $<1.7$ & $<0.9$\\ 
K6 & 2.0 / 4.0 / 1.6 & $\sim 30$  & $<0.1$ & $<0.2$ & $<3.7$ & $<1.9$ & $<1.6$ & $<1.0$\\
K8 & 1.7 / 2.3 / 1.7 & $\sim 30$  & $<0.1$ & $<0.1$ & $<3.2$ & $<1.1$ & $<1.4$ & $<0.6$\\
\hline
\end{tabular}
\tablefoot{
The central velocity of the line in the B1 position is 103~km\,s$^{-1}$. 
Col.\,(1): Source name. Col.\,(2): Noise levels in WILMA CO(1-0), WILMA 
CO(2-1), and 4MHZ spectra with 10.4~km\,s$^{-1}$ channels. From fitting 
first-order baselines in a 2000~km\,s$^{-1}$ wide region of the 
spectra. Col.\,(3): Linewidths used for both CO(1-0) and CO(2-1) lines. 
Measured in B1 from the CO(2-1) detection, estimated for other 
positions (see the text). Col.\,(4): Detected (B1) or $3\sigma$ upper 
limits on WILMA CO(1-0) integrated intensity. Col.\,(5): Detected (B1) 
or $3\sigma$ upper limits on CO(2-1) integrated intensity. Col.\,(6): 
Detected (B1) or $3\sigma$ upper limits on CO(1-0) luminosity. 
Col.\,(7): Detected (B1) or $3\sigma$ upper limits on CO(2-1) 
luminosity. Col.\,(8): Detected (B1) or $3\sigma$ upper limits on 
molecular gas mass, including a factor of 1.36 to account for the 
effect of He, corresponding to WILMA CO(1-0) luminosity. 
Col.\,(9): Detected (B1) or $3\sigma$ upper limits on molecular gas 
mass, including a factor of 1.36 to account for the effect of He, 
corresponding to WILMA CO(2-1) luminosity.
} 
\end{table*}

\begin{figure*}[t]
\centering
\includegraphics[height=0.33\textwidth,angle=270]{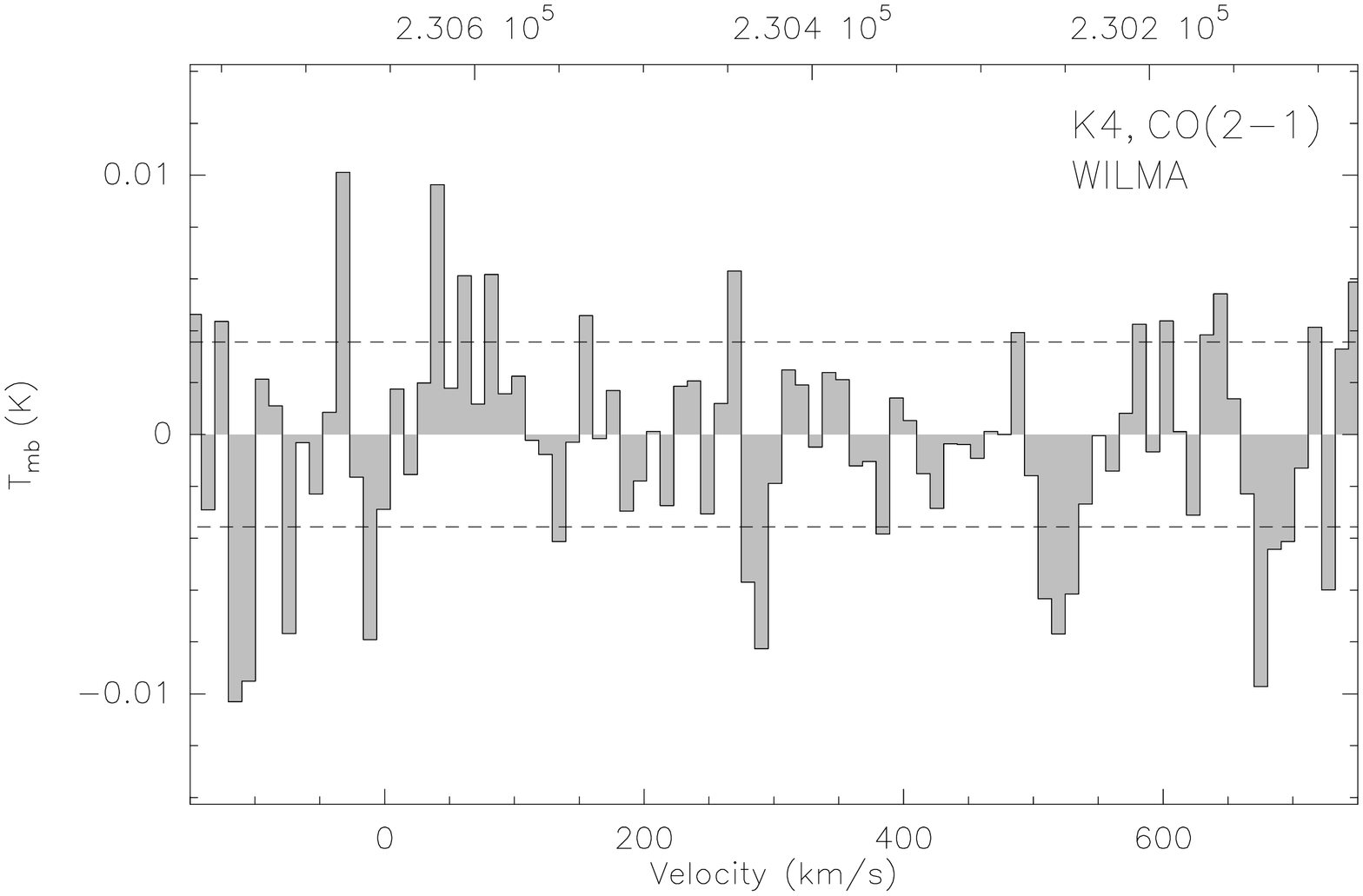}
\includegraphics[height=0.33\textwidth,angle=270]{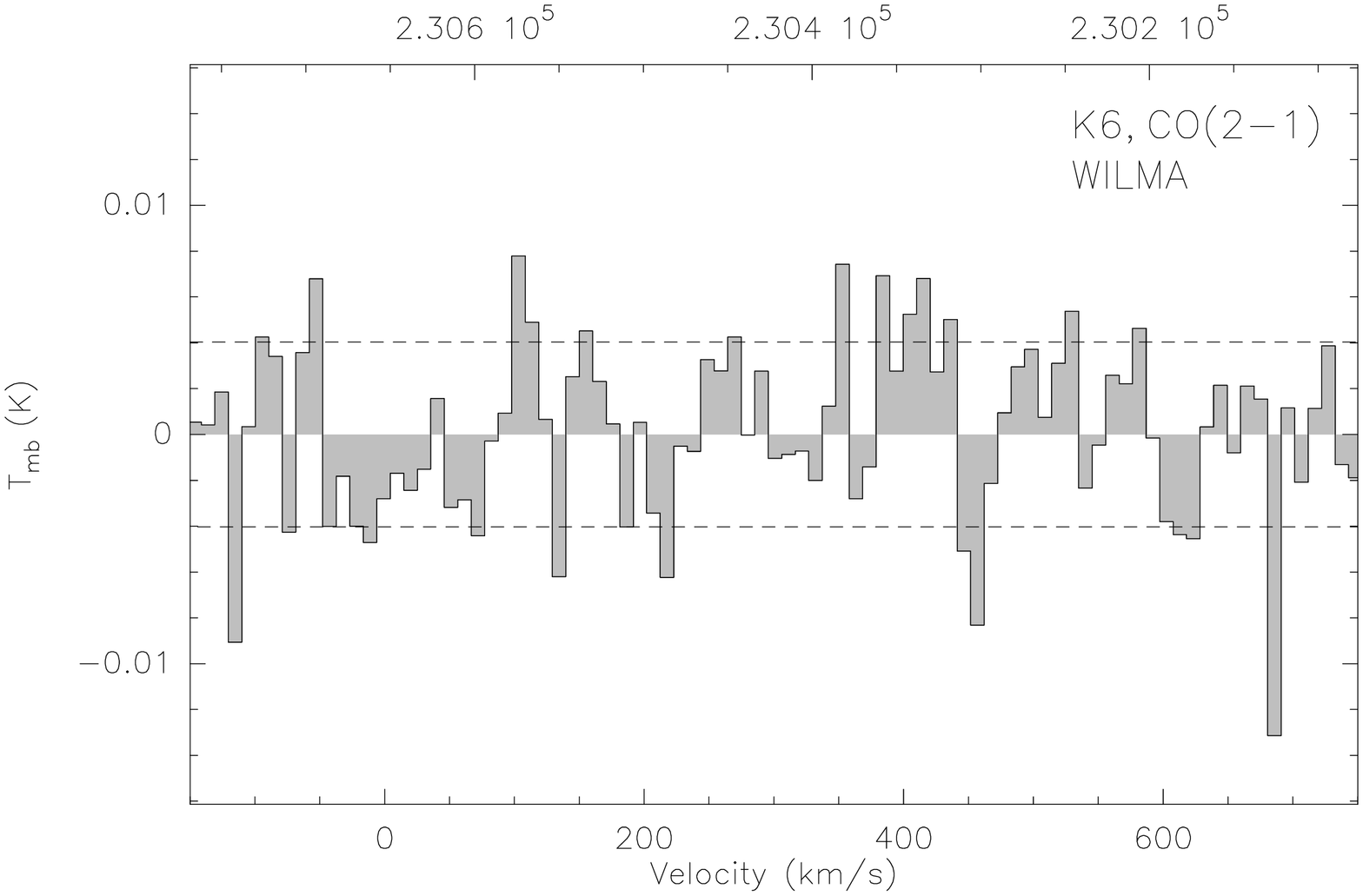}
\includegraphics[height=0.33\textwidth,angle=270]{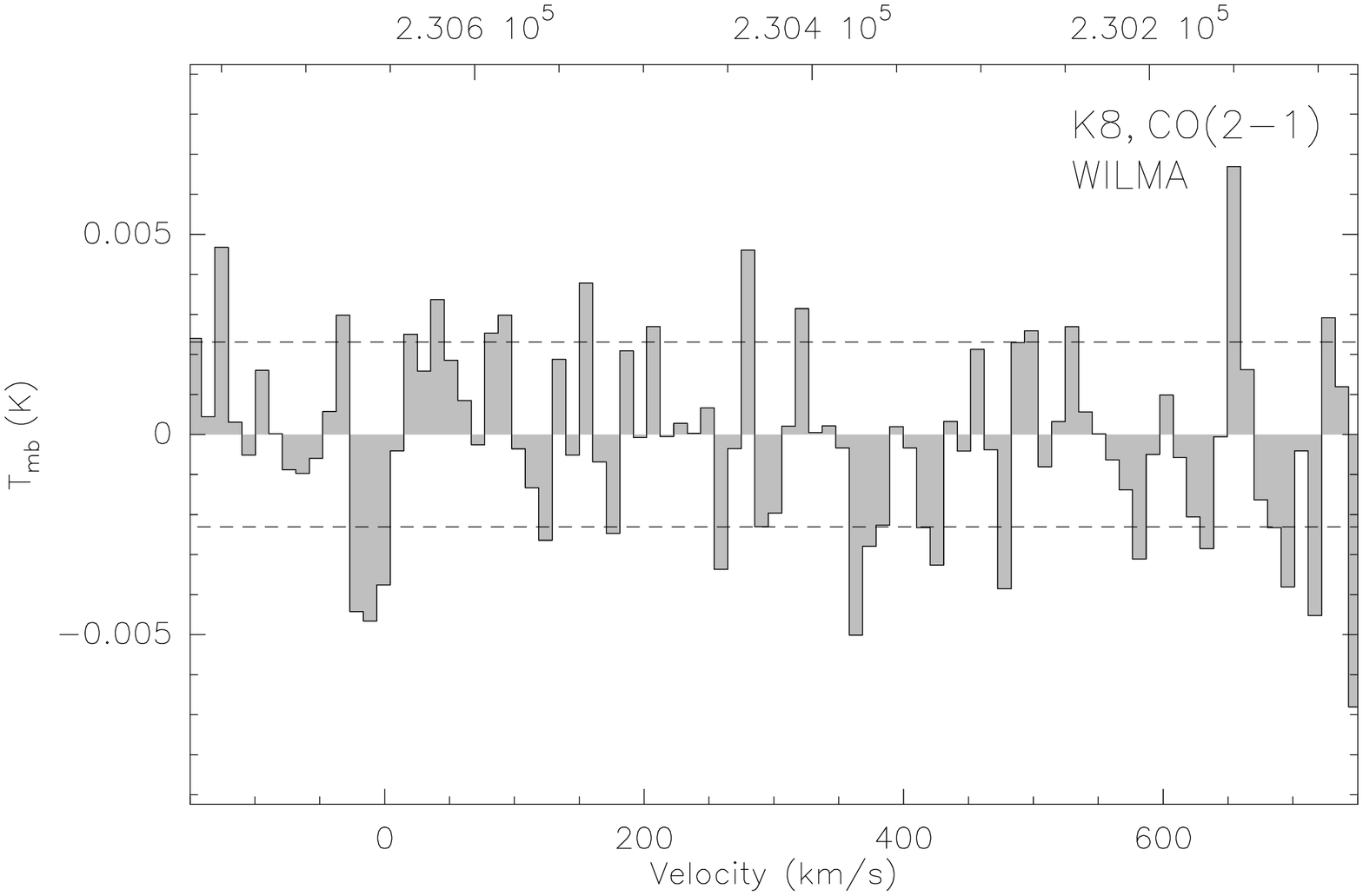}
\includegraphics[height=0.33\textwidth,angle=270]{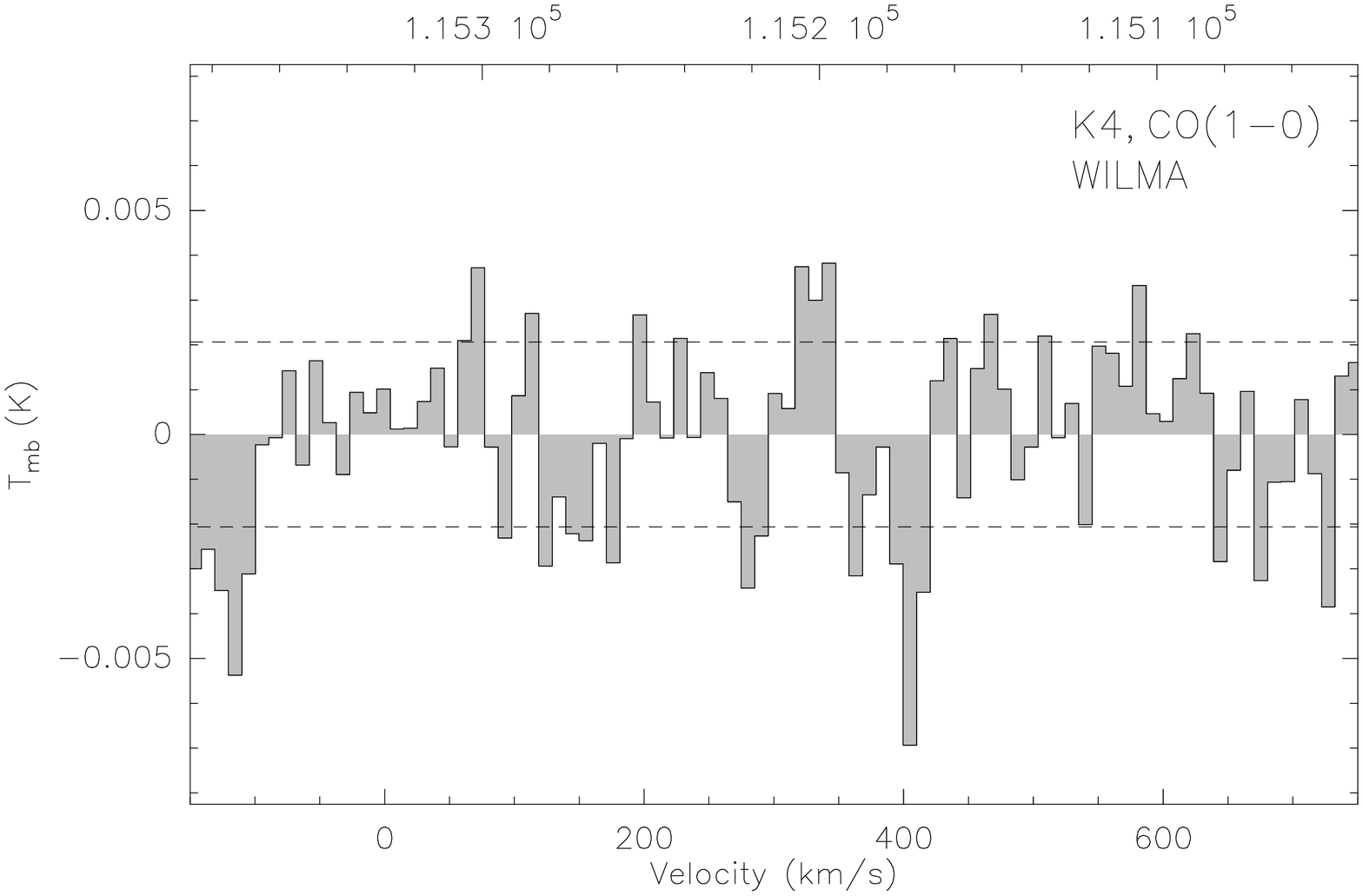}
\includegraphics[height=0.33\textwidth,angle=270]{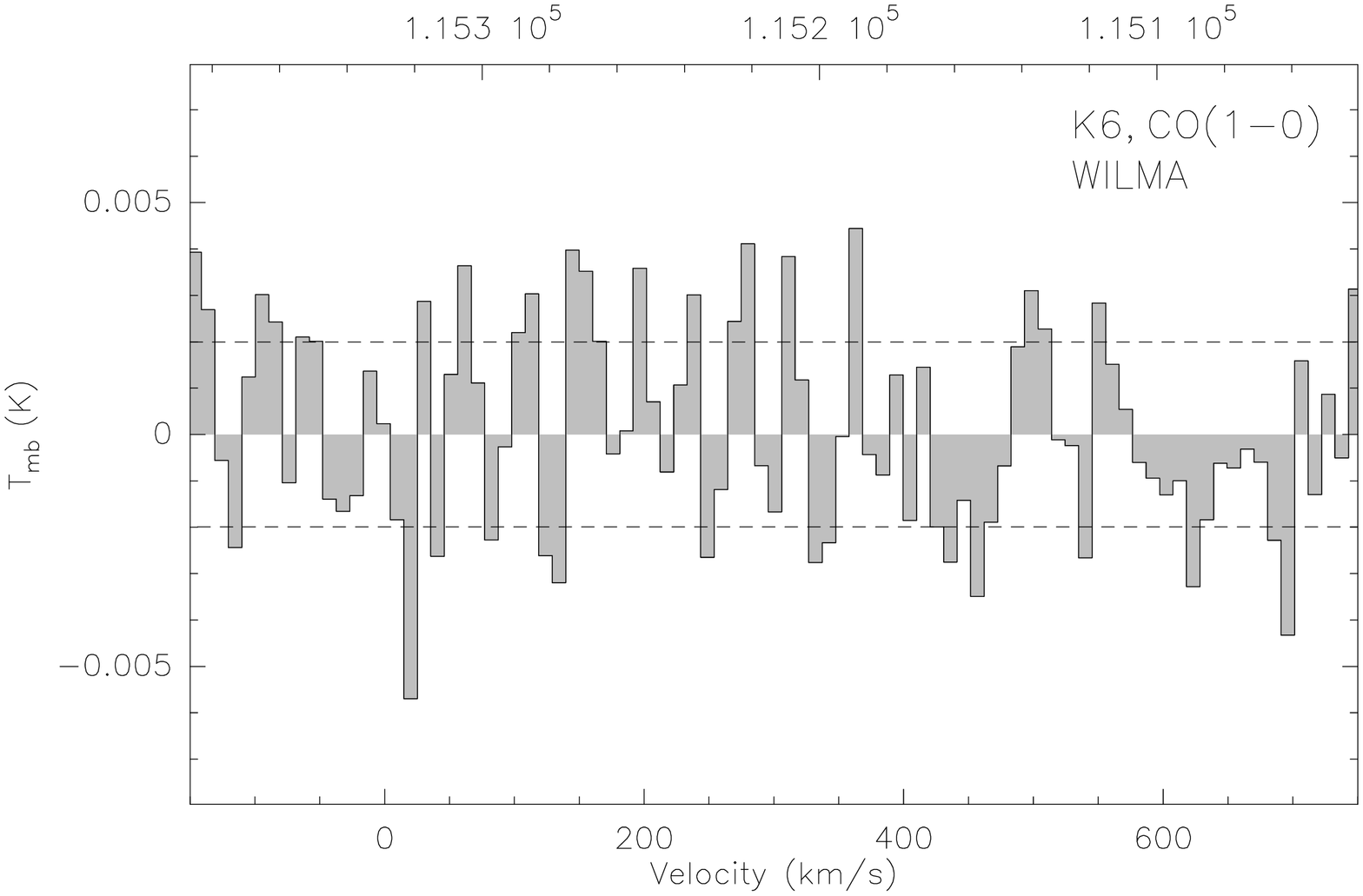}
\includegraphics[height=0.33\textwidth,angle=270]{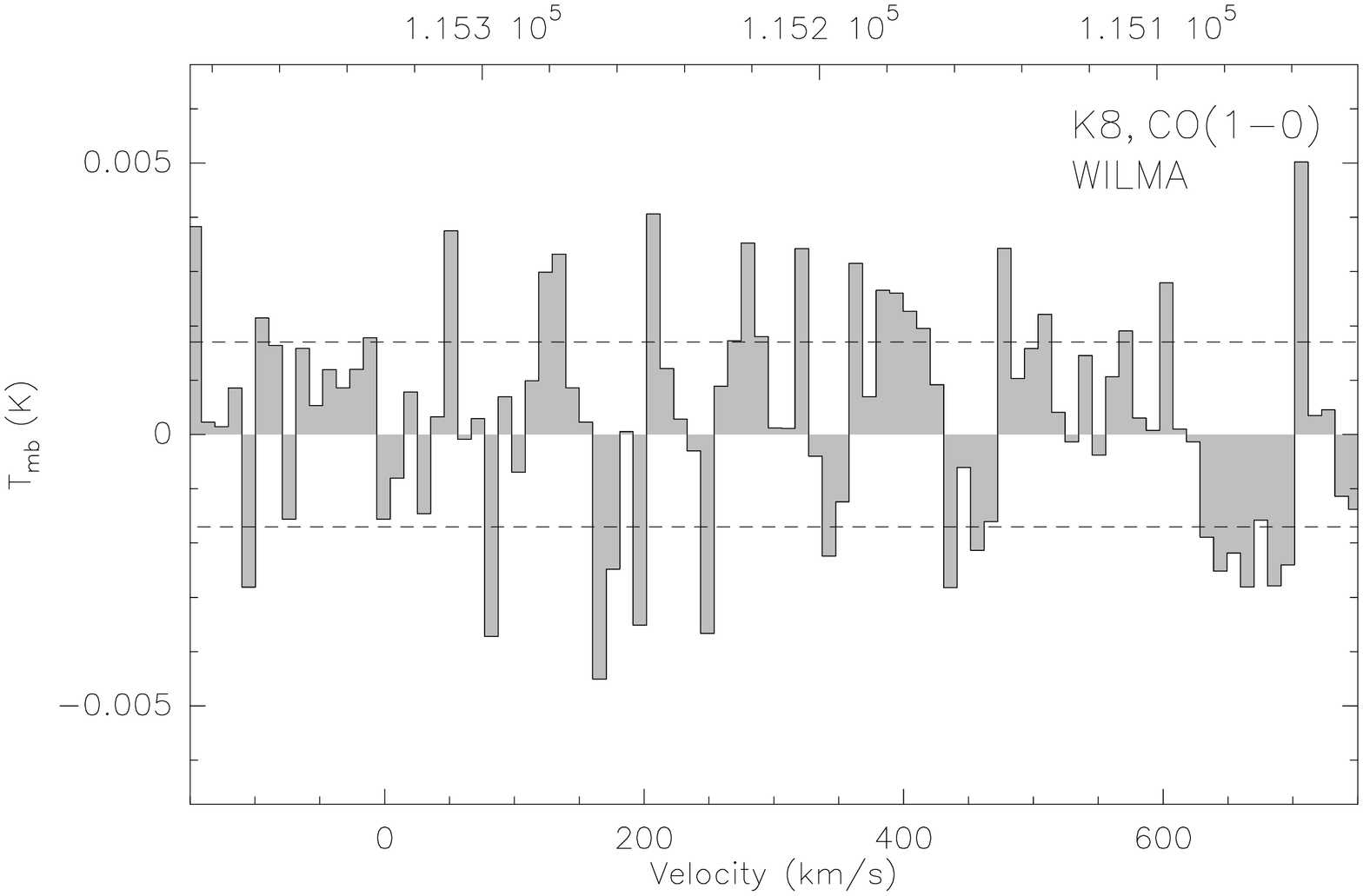}
\caption{
Tail positions K4, K6, and K8: WILMA CO(2-1) ({\it top}) and WILMA 
CO(1-0) ({\it bottom}) spectra smoothed to 10.4~km\,s$^{-1}$ 
resolution. Dashed lines show $\pm 1\sigma$ noise levels; $y$-axis 
scaling is set to $\pm 4\sigma$. 
}\label{FigK6K8K4}
\end{figure*}

There is no statistically significant line detection in the CO(1-0) 
WILMA and 4MHz spectra of the central position, but there are 
$2.0-2.7$-$\sigma$ features with the same velocity range as the CO(2-1) 
line. The integrated CO(1-0) intensity measured from the WILMA spectrum 
is $\sim 0.06$~K\,km\,s$^{-1}$ (see Table~\ref{Results}). The CO(1-0) 
and CO(2-1) features are consistent with a compact CO source which 
suffers much more beam dilution in the $21\arcsec$ CO(1-0) beam than the 
$11\arcsec$ CO(2-1) beam. With four times the beam area, a point-like source 
will experience four times the beam dilution in the CO(1-0) beam. The 
ratio of peak temperatures of $\sim 3$ and integrated intensities of 
$\sim 4$ are consistent with the factor of 4 expected from maximum beam 
dilution multiplied by the typical $I_{\rm CO(2-1)}/I_{\rm CO(1-0)}$ 
line ratio of $0.8-1$ in molecular clouds \citep[e.g.,][]{leroy2008}. 

We thus consider the CO(2-1) emission in the center of IC3418 as 
marginally detected. In Fig.~\ref{FigB1B2K}, the CO(2-1) spectral line 
is fitted with a Gaussian profile with the central velocity of 
103~km\,s$^{-1}$, FWHM of 54~km\,s$^{-1}$, peak brightness temperature 
of 4.3~mK, and the integrated intensity $I_{\rm CO(2-1)}= 0.24\pm 
0.05$~K\,km\,s$^{-1}$. The central velocity of the line is somewhat 
smaller than that deduced by Kenney et al. (in prep.) from the stellar 
Keck spectra (see Table~\ref{ic3418}). We discuss the velocity 
difference in Sect.~\ref{centralbulk}. Since there is no \ion{H}{i} emission 
in IC3418, we estimate the maximum rotation velocity from the 
Tully-Fisher relation (by comparing the $H$-band luminosity of IC3418 
to other Virgo galaxies with known rotation velocities) to be about 
55--60~km\,s$^{-1}$. Correcting for an inclination of $\sim 50\degr$ 
\citep{chung2009}, the expected linewidth is about 50~km\,s$^{-1}$, 
similar to the observed linewidth. 

At the adopted Virgo distance of 16.5~Mpc, the main beam projected area 
at CO(2-1) frequency is $\Omega_B \simeq 137$~arcsec$^2= 0.9$~kpc$^2$, 
including a Gaussian beamshape correction factor of $1/\ln2$. The B1 
integrated intensity then corresponds to a luminosity $L_{\rm CO(2-1)}= 
2.1\times 10^5$~K\,km\,s$^{-1}$\,pc$^2$. The corresponding molecular 
gas mass can be calculated using $M_{\rm H_2}\, [M_\odot]= 5.5\, X_{\rm 
CO}\, \frac{R_{21}}{0.8}\, L_{\rm CO}\, [{\rm K\,km\,s^{-1}\,pc^2}]$, 
where $X_{\rm CO}= N_{\rm H_2}/ I_{\rm CO}$ is the CO(1-0)-to-H$_2$ 
conversion factor normalized to a standard Galactic value of $2\times 
10^{20}$~cm$^{-2}$(K\,km\,s$^{-1}$)$^{-1}$ \citep[e.g., ][]{kennicutt2012, 
pineda2010, feldmann2012II}, and $R_{21}$ is the CO(2-1) to CO(1-0) 
ratio. We assume a typical value of $R_{21}=0.8$ \citep[e.g., 
][]{leroy2008}. The above formula includes a factor of 1.36 to account 
for the effects of helium. The resulting amount of molecular gas 
possibly detected in the B1 main body position is $1.2\times 
10^6~M_\odot$. In Sect.~\ref{Xfactor} we discuss that the X-factor may 
be (somewhat) larger than the standard Galactic value because of the 
sub-solar metallicity of IC3418. 

The size of the emitting region in the B1 pointing is certainly 
comparable to or smaller than the CO(2-1) beam, i.e., $\la 900$~pc. 
This is most likely gas concentrated in the nuclear region. A lower 
limit on the column density of the central CO-emitting blob is $\propto 
M_{\rm mol}/ \Omega_B\sim 1~M_\odot$\,pc$^{-2}$, assuming a standard 
Galactic conversion X-factor, if gas uniformly fills the beam. It is, 
however, much more likely to be concentrated over a smaller area. The 
actual gas surface density could thus be similar to that of a typical 
giant molecular cloud (GMC), or $\sim 100~M_\odot$\,pc$^{-2}$. Also 
the B1 luminosity of $2.1\times 10^5$~K\,km\,s$^{-1}$\,pc$^2$ is 
comparable to the luminosity of a single giant Galactic molecular 
cloud. However, the linewidth is much larger than that from a single 
GMC with $M_{\rm gas}\simeq 10^6~M_\odot$ (typically $\sim 
10$~km\,s$^{-1}$) which strongly suggests that emission is not from a 
single GMC. 

In the off-center main body position B2 (about $15\arcsec\sim 1.2$~kpc from 
B1) covering the NE young stellar complex (which is the brightest 
optical part of the galaxy), no CO emission was detected (see 
Fig.~\ref{FigB1B2K}, middle panels). This suggests that 
molecular gas has possibly survived the effects of the cluster 
environment only in the very center of the galaxy. In the B2 position, 
we assume an upper limit on the line width to be about 30~km\,s$^{-1}$, 
i.e., smaller than at B1, as it does not cover the rotation center. 
Rotation curve studies of Local Volume dwarfs 
\citep[e.g.,][]{kirby2012} show that at $r=1$~kpc, the rotation 
velocity is $0.5 \upsilon_{\rm max}\sim 30$~km\,s$^{-1}$. Thus, we 
derive the upper limits on the integrated CO intensity, CO luminosity, 
and molecular gas mass, given in Table~\ref{Results}. 

\subsection{Stripped tail and fireballs}\label{iramtail}
We searched for CO emission in the three points K4, K6, and K8 
covering most of the bright \ion{H}{ii} regions in the outer tail (see 
Fig.~\ref{Figiram}). The positions K4 and K6 show a head-tail (fireball) 
structure with H$\alpha$ offset from the UV peaks. We didn't observe 
the brightest UV region \citep[K5 in the notation of][]{hester2010} and 
instead went for the more distant knot K8 that has a higher H$\alpha$/UV 
ratio than K5, suggesting that it may be at an earlier evolutionary 
stage with more gas. The galaxy IC3418 is moving toward us with respect 
to M87 and the stripped material is thus extending away from us: we 
expect it to be accelerated to larger line-of-sight velocities. 

In Fig.~\ref{FigK6K8K4} we inspect the WILMA spectra for the K4, K6, 
and K8 regions over a large range, from velocities of the main body to 
those several hundred km\,s$^{-1}$ higher. Although there are few $\sim 
3\sigma$ features resembling spectral lines, they never occur in the 
spectra at both frequencies simultaneously, as in the case of B1. As we 
will show later in Figs.~\ref{FigSigISM-prepeakVz} and 
\ref{FigSigISM-postpeakVz}, only low-column density ISM parcels could be 
accelerated by ram pressure to large vertical velocities. No clear line 
emission is thus detected in any of the three tail positions. Since the 
tail linewidths might be broadened by galactic rotation, we will assume 
linewidths of $\sim 30$~km\,s$^{-1}$ to derive upper limits on CO 
intensity in all three tail positions. Table~\ref{Results} gives 
corresponding upper limits on $I_{\rm CO}$, $L_{\rm CO}$, and $M_{\rm 
mol}$. The upper limits on the molecular gas mass are $\sim 1\times 
10^6~M_\odot$ per point, assuming a standard Galactic X-factor, but it 
is possible that in the H$\alpha$ bright tail regions the CO-to-H$_2$ 
conversion factor is different from standard value (see discussion in 
Sect.~\ref{Xfactor}). 

In order to increase the SNR, we also combined data from the three 
positions, assuming for a moment that conditions of the molecular gas 
are similar and that their radial velocities are close; however, no 
significant signal occurred in the stacked spectrum (see the right 
panels in Fig.~\ref{FigB1B2K}). Our achieved upper limits per point on 
CO luminosity are comparable to the luminosity across most of the SMC, 
$L_{\rm SMC, CO10} \sim 1\times 10^5$~K~km~s$^{-1}$~pc$^2$ as measured 
by \citet{mizuno2001}. With CO(1-0) point luminosities of $\sim 2\times 
10^5$~K\,km\,s$^{-1}$\,pc$^2$, we are sensitive enough to detect 
analogues to the brightest clouds in M33 or the LMC N197 complex. 

\section{Current molecular gas content}\label{current}
In this section we will estimate from our IRAM observations upper 
limits on the molecular content in the whole galaxy. We will also 
calculate an upper limit on the molecular gas mass following from dust 
observations of IC3418. 

\subsection{Current upper limits -- the whole galaxy}
In Table~\ref{limits} we give estimates (upper limits) of the CO 
luminosity and molecular gas mass in the central beam of the main body 
(the B1 position), the entire main body, the tail, and the whole galaxy 
(main body plus tail). To estimate the total CO luminosity of the main 
body of the galaxy from the measured B1 and B2 (upper limit) 
luminosities, we estimate the filling factor of the CO(2-1) and CO(1-0) 
beams as the ratio of their area to the optical surface of the galaxy, 
assuming an exponential distribution of CO emission with a scale length 
roughly 1.5 times smaller than the optical scale length \citep{young1995}. 
\citet{fumagalli2011} fitted the surface brightness of the IC3418 disk 
with an exponential law with a scale length of $19\arcsec\simeq 1.5$~kpc. 
Accounting for the $R_{25}$ major axis diameter of $\sim 7.2$~kpc and 
the inclination of $\sim 50\degr$ (see Table~\ref{ic3418}), we 
estimate that the CO(1-0) main beam located in the B1 position would 
encompass about $25\%$ of the total disk CO luminosity. This yields an 
estimated upper limit on the molecular gas content in the whole galaxy 
of about $5\times 10^6~M_\odot$. \citet{leroy2008} suggested that 
typical dwarfs in a sample of nearby galaxies have molecular gas 
distributed mostly in their inner ($<0.25 r_{25}$) radius, which 
matches the CO(1-0) beam radius. Moreover, since the galaxy is probably 
gas-stripped from the outside in, the expected gas distribution would 
be more compact than the stellar disk. Thus, it is likely that we are 
detecting a larger fraction of the total CO flux in the central beam. 

Applying our $3\sigma$ upper limit of $M_{\rm mol}\la 1\times 
10^6~M_\odot$ measured in each of the three tail regions (K4, K6, and K8) to 
the total number of knots occurring in the wake, an upper limit on the 
total molecular mass in the tail would be $\sim 9\times 10^6~M_\odot$. 
A similar value comes from calculating the filling factor of the three 
CO(1-0) beams in the whole tail area. Assuming that the molecular gas 
occurs (mostly) in the actual places of star formation, and taking into 
account that all \ion{H}{ii} regions are observed only in the outer 
half of the tail, the upper limit could possibly be lowered to a value 
similar to the one derived above for the main body. Nevertheless, we 
conclude that the upper limit on the current molecular gas content of 
the whole galaxy (main body plus tail) is about $1.5\times 10^7~M_\odot$. 

\begin{table}
\centering
\caption[]
{Molecular gas mass estimates corresponding to our IRAM 
detection and upper limits on CO luminosity in the main body and in the 
tail, assuming a Galactic CO-to-H$_2$ conversion factor.}
\label{limits}
\begin{tabular}{lccc}
\hline\hline
\noalign{\smallskip}
Source & $L_{\rm CO(2-1)}$		   & $M_{\rm mol}$\\
       & ($10^5$~K\,km\,s$^{-1}$\,pc$^2$) & ($10^6~M_\odot$)\\
\noalign{\smallskip}
\hline
\noalign{\smallskip}
center (B1) & 2.1 & 1.2\\
total disk  & $\geq2.1$, $<9$ & $\geq1.2$, $<5$\\
total tail  & $<18$ & $<10$\\
body + tail & $<27$ & $<15$\\
\hline
\end{tabular}
\end{table}

\subsection{Current molecular gas amount from dust emission}
Far infrared (FIR) dust measurements allow an alternative approach for 
estimating the current molecular gas mass that overcomes some 
disadvantages of traditional CO line tracers. It traces the total gas 
column density (\ion{H}{i}+H$_2$). In the Small Magellanic Cloud, IRAS 
observations suggested much more H$_2$ than seen from CO 
\citep{israel1997, leroybolatto2007} and, thus, that CO may be 
underabundant or absent in regions where H$_2$ survives because the 
H$_2$ self-shields while the CO is photo-dissociated 
\citep[e.g.,][]{maloney1988}. 

From ISO observations \citet{tuffs2002} obtained $3\sigma$ upper limits 
on 60~$\mu$m, 100~$\mu$m, and 170~$\mu$m FIR dust emission of IC3418 of 
$<0.04$~Jy, $<0.03$~Jy, and $<0.07$~Jy, respectively. The total dust 
mass can be determined to within a factor of 3 from 
\citep[e.g.,][]{boselli2002, evans2005} 
\begin{equation}
M_{\rm dust}= C\,S_{100\,\mu{\rm m}}\, D^2\, ({\rm e}^{a/T_{\rm 
dust}}-1),
\end{equation}
where $C$ relates to the grain opacity, $S_{100\,\mu{\rm m}}$ is the 
flux in the 100~$\mu$m band in Jy, $D$ is the distance of the galaxy in 
Mpc, and $T_{\rm dust}$ is the dust temperature. According to 
\citet{boselli2002}, we use $C=1.27~M_\odot$\,Jy$^{-1}$\,Mpc$^{-2}$,  
$T_{\rm dust}= 20.8$~K, and $a=144$~K for 100~$\mu$m emission. For the 
$S_{100\,\mu{\rm m}}< 0.03$~Jy upper limit we get a limit on the dust 
mass of $\la 10^4~M_\odot$. Taking a typical cold gas-to-dust 
ratio of about 500 \citep[e.g.,][]{lisenfeld1998}, we obtain an upper 
limit on the molecular gas mass in the galaxy of $\la 5\times 
10^6~M_\odot$. This is a value similar to our CO luminosity-based upper 
limit estimate (see Table~\ref{limits}). As we will discuss in the 
following section, it is likely that molecular gas has been effectively 
stripped from IC3418. This would imply that some dust also has been 
removed from the galaxy by ram pressure stripping. 

\section{Original molecular gas content}\label{original}
We estimate the molecular gas mass $M_{\rm mol}$ likely to have been in 
IC3418 before it was ram pressure stripped from the star formation rate 
(SFR) in the main body. The FUV emission is a tracer of star formation 
in the last couple 100~Myr. The FUV luminosity within the 
25~mag\,arcsec$^{-2}$ $B$-band isophote of the main body of IC3418 
(corrected for Galactic extinction) is $L_{\rm FUV}= 1.4\times 
10^{41}$~erg\,s$^{-1}$ \citep{gildepaz2007}. The classical 
SFR--luminosity relation works well for continuous star formation 
approximation where SFR is constant over the timescale of the UV 
emission. This is probably not true in IC3418, where star formation 
ceased about 200~Myr ago as a result of ram pressure stripping of 
star-forming ISM, and so the current FUV luminosity is likely scaled 
down because the stellar population had faded. From evolutionary 
stellar population modeling we can estimate that FUV flux in a case 
where star formation ended 100~Myr or 200~Myr ago is reduced by a 
factor of 8 or 20, respectively, compared to a model with continuous 
star formation (based on private communication with Hugh Crowl). This 
factor could be somewhat lower if ram pressure induced a burst of 
star formation before quenching it, a phenomenon that was identified in 
the outer radii of NGC~4522 \citep{crowl2006}, for example. For IC3418 
this is suggested from measuring equivalent widths of the absorption 
Balmer lines in the spectrum of IC3418, which are stronger than in post 
star-burst k+a galaxies \citep{fumagalli2011}. 

As a result, we will apply a factor of $\sim$10 to the star formation 
rate of $6.1\times 10^{-3}~M_\odot$\,yr$^{-1}$ determined from 
the FUV luminosity--SFR formula of \citet{kennicutt2012} 
\begin{equation}\label{kennicutt}
\log{\rm SFR}\ (M_\odot\,{\rm yr}^{-1})= \log{L_{\rm FUV}}\ ({\rm 
erg\,s^{-1}}) - 43.35.
\end{equation}
To account for sub-solar metallicities appropriate for dwarf galaxies, 
we will also apply to the SFR the factor of $1/1.1$ estimated by 
\citet{hunter2010} from the stellar population evolution models of 
STARBURST99 \citep{leitherer1999} for luminosity at $1500~\AA$. This 
then yields an original star formation rate of $\sim 5.6\times 
10^{-2}~M_\odot$\,yr$^{-1}$ in IC3418, a value that is consistent with 
measurements in other dwarf galaxies where SFRs can span several orders 
of magnitude \citep[e.g.,][]{hunter2010, schruba2012}. Assuming a 
typical star formation efficiency in dwarf galaxies of $\tau_{\rm 
dep}\approx 2$~Gyr \citep[e.g.,][]{bigiel2008}, this yields about 
$1\times 10^8~M_\odot$ of molecular gas that originally was in the 
galaxy. This is about 20~times more than our CO-based upper limit 
estimate of the current molecular gas content of the main body. 

Another rough estimate of the original molecular amount can come from 
the stellar mass of the galaxy and typical fractions of atomic and 
molecular gas seen in galaxies of the same type. However, this estimate 
is affected by the issue of the X-factor that we will deal with later. 
\citet{bothwell2009} calculated the stellar mass-to-light ratios for 
different galaxy morphology-color combinations using the algorithm of 
\citet{belldejong2001}. For Im and dIrr types they derived $M/L_B= 
0.49$. For the $B$-band magnitude $m_B= 14.85$ (see Table~\ref{ic3418}) 
of IC3418 this yields a stellar mass $M_*\approx 3\times 10^8~M_\odot$, 
which is consistent with the estimate of \citet{fumagalli2011} ($\sim 
3.8\times 10^8~M_\odot$). It is known that ISMs of dwarf galaxies are 
dominated by large reservoirs of atomic gas -- along the Hubble 
sequence the \ion{H}{i} fraction increases up to typically $M_{\rm 
\ion{H}{i}}/M_*>0.5$ 
or even $>1$ in the late types \citep[e.g.,][]{belldejong2000}. The 
original \ion{H}{i} mass of IC3418 is estimated to $\sim 6\times 10^8~M_\odot$ 
\citep{gavazzi2005}, which is in agreement with the expected $M_{\rm 
\ion{H}{i}}/M_*>1$. Although dwarf galaxies are \ion{H}{i}-rich, molecular gas 
represents only a small fraction of their total gas mass, typically 
$M_{\rm H_2}/M_{\rm \ion{H}{i}}\approx 0.1-0.2$ which is similar to the ratio 
found in the outer parts of spiral galaxies 
\citep[e.g.,][]{obreschkow2009, israel1997}. This corresponds to $\sim 
(0.6-1)\times 10^8~M_\odot$ of original H$_2$ in IC3418, which is 
consistent with the above estimate. 

\subsection{Comparison with other galaxies}
\begin{figure}
\centering
\includegraphics[height=0.48\textwidth,angle=270]{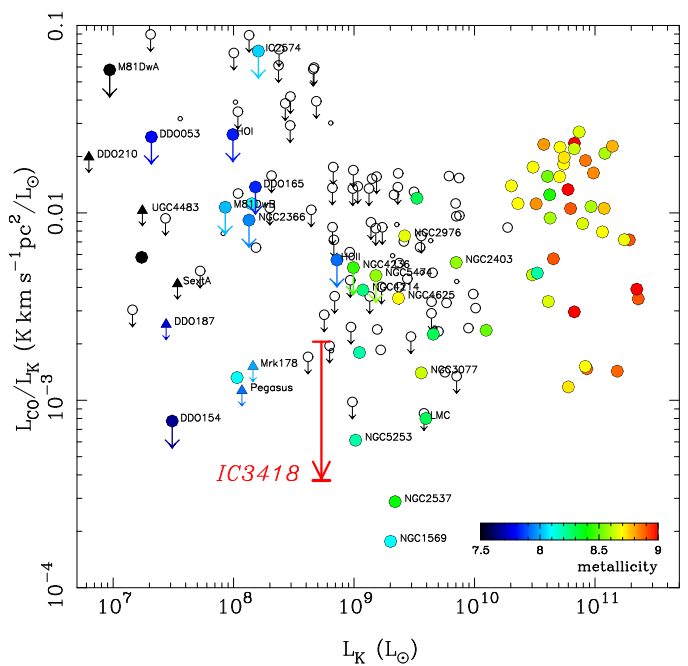}
\caption{
Normalized CO(1-0) luminosity as a function of $K$-band luminosity. 
IC3418 is represented with a red arrow with the head corresponding to 
the marginally detected (B1) CO luminosity and the upper tail bar to 
the upper limit for the total luminosity of the disk. IC3418 is 
compared with a set of nearby low-mass star-forming galaxies from the 
HERACLES survey (together with more massive HERACLES galaxies, and 
other Local Group and nearby galaxies) used by \citet{schruba2012} 
({\it filled circles}), and with a sample of compact late-type spiral 
and irregular galaxies of \citet{leroy2005} ({\it empty circles}), plus 
few starburst dwarf galaxies from \citet{taylor1998} ({\it triangles}), 
and DDO154 \citep{komugi2011}. Upper limits are plotted with symbols 
with small arrows. Colors represent metallicity $12+\log{(O/H)}$. The 
color of the IC3418 arrow does not correspond to its metallicity (see 
Sect.~\ref{Xfactor}). Small empty circles show marginal detections of 
\citet{leroy2005}. 
}\label{FigLCOLK}
\end{figure}

We will also compare our results with other galaxies, including 
low-mass, low-metallicity dwarfs. In Fig.~\ref{FigLCOLK}, IC3418 is 
placed into a plot of normalized CO luminosity $L_{\rm CO}/L_K$ vs. 
$L_K$, together with (a) the sample of \citet{schruba2012} containing 
16 nearby low-mass star-forming HERACLES galaxies as well as some 
more massive HERACLES galaxies and Local Group galaxies, (b) a subset 
of the sample of compact late-type spirals and irregulars of 
\citet{leroy2005} (their detections correspond instead to lower limits 
because they observed only the central few kpc of each galaxy), (c) a 
few starburst dwarfs of \citet{taylor1998} not contained in the 
\citet{schruba2012} collection, and (d) DDO154 observed by 
\citet{komugi2011}. 

The normalization $L_{\rm CO}/L_K$ used in Fig.~\ref{FigLCOLK} should 
remove most of the galactic mass and size correlations and thus reveal 
an unbiased state of the CO content of the galaxies. All the $K$-band 
magnitudes of the galaxies were taken from NED, as well as their 
distances \citep[except for the sample of][for which we used their 
distances]{schruba2012}. We preferred to use redshift-independent 
(mostly Tully-Fisher) distances whenever available. The $K$-band 
apparent magnitude of IC3418 is 12.55 (see Table~\ref{ic3418}). The 
CO(2-1) luminosity detected in the B1 position was converted to CO(1-0) 
scale using $I_{\rm CO21} = 0.7\,I_{\rm CO10}$, to be consistent with 
\citet{schruba2012}. 

In Fig.~\ref{FigLCOLK} IC3418 is represented with a red arrow with the 
head corresponding to the detected (B1) CO luminosity and the upper 
tail bar to the upper limit for the total luminosity of the disk. When 
first focusing on the high end of the IC3418 range, the galaxy occurs 
in the middle of other galaxies and thus shows no special value of 
$L_{\rm CO}/L_K$. However, when taking into account only the actual 
detection in the nuclear region of the disk, IC3418 shifts to a rather 
sparsely populated region of the plot where only few galaxies have 
similar or lower values of the normalized CO luminosities. Although 
this region may suffer from a lack of observations with CO sensitivity 
levels similar to ours, this suggests that CO content of IC3418 is low 
relative to its other properties. The $L_{\rm CO}/L_K$ ratio is 
systematically 
lower in galaxies with $L_K\la 10^{10}~L_\odot$, probably mostly 
because of a different CO-to-H$_2$ relation connected to their low 
metallicity. But even among low-mass galaxies of a given metallicity, 
the CO-to-H$_2$ relationship may vary due to variations in the amount 
of photo-dissociating UV radiation, which is correlated with the star 
formation rate. The other dwarfs with similar or lower values of 
$L_{\rm CO}/L_K$, notably NGC~1569, NGC~2537, or NGC~5253, are all 
starburst blue compact dwarfs in which strong photo-dissociation of 
molecular gas by UV radiation from young stars may make their CO 
content especially low. In IC3418, since star formation ended about 
$100-200$~Myr ago, the effects of photo-dissociation are expected 
to be less extreme. 

Thus, the reason for the low $L_{\rm CO}/L_K$ ratio of IC3418 must be 
different. Since the galaxy clearly has been strongly ram pressure 
stripped, the removal of molecular or molecular-forming ISM would be a 
possible explanation. 

\subsection{CO as a tracer of H$_2$}\label{Xfactor}
It has been shown that in dwarf irregular galaxies with metallicities 
lower than about $1/10$ of solar, i.e., $12+\log(O/H)\leq 8.0$, the 
detection of CO emission steeply decreases \citep[e.g.,][]{taylor1998, 
leroy2007, komugi2011, schruba2012}, although ionized carbon and the 
ratio of IR to millimeter dust emission around star-forming regions 
indicate that molecular gas is still present. Applying a Galactic 
CO-to-H$_2$ conversion ratio then results in very low molecular gas 
masses and about $10-100$ times higher star formation efficiencies 
(SFEs). It is, however, more likely that the true SFEs of dwarfs are 
normal, and the CO-to-H$_2$ factor is systematically different. Thus, 
assuming a constant H$_2$ depletion time of $\tau_{\rm dep}=1.8$~Gyr, 
\citet{schruba2012} derived for dwarf galaxies with metallicities $1/2 
- 1/10~Z_\odot$ conversion factors more than one order of magnitude 
larger than in massive spirals with solar metallicities. 

Using evolutionary models for Lick indices in Keck spectroscopic data, 
stellar metallicity that represents an average over the star formation 
history in IC3418 is estimated to $0.5\pm 0.2$ solar, i.e., $12+ 
\log(O/H)\sim 8.4$ (Kenney et al., in prep.)\footnote{We assume a solar 
value of $12+ \log(O/H)= 8.69$ \citep{asplund2009}}. Another 
estimate comes from the stellar mass-metallicity relation 
\citep{tremonti2004, kewley2008, pettini2004}: For the IC3418 stellar 
mass of $\sim 3\times 10^8~M_\odot$, the predicted stellar metallicity 
is $12+ \log(O/H)\sim 8.3$, i.e., $\sim 0.4$ solar. Unfortunately, 
there is no data available on the gas metallicity of IC3418 that would 
reveal the actual amount of metals currently present in the 
interstellar medium. Nevertheless, in IC3418 we expect the CO-to-H$_2$ 
conversion factor to be larger than Galactic because of the decreased 
dust shielding at a presumably low metallicity in the galaxy. From 
Eq.~5 of \citet{leroy2013}\footnote{
$c_{\rm CO, dark}\approx 0.65 \exp{[0.4/(D/G)']}$, where $(D/G)'$ is 
the dust-to-gas ratio normalized to the Galactic value of 0.01 (see 
Sect.~\ref{current}). A typical surface density of a GMC of 
$100~M_\odot$\,pc$^{-2}$ is adopted.} the correction to the standard 
CO-to-H$_2$ conversion factor can be estimated to $\sim 5$. However, 
this estimate remains highly uncertain. 

\subsection{H$_2$-deficiency?}
While many Virgo cluster spirals are highly \ion{H}{i}-deficient 
\citep[e.g.,][]{kenney2004, chung2007, chung2009, abramson2011}, their 
molecular gas content is relatively normal or only slightly deficient 
\citep[e.g.,][]{kenneyyoung1986, kenneyyoung1989}. This is largely a 
difference between the inner H$_2$-dominated, and outer \ion{H}{i}-dominated 
disks. In galaxies which are clearly ram pressure stripped, nearly all 
the gas, including molecular gas, and dust beyond some gas truncation 
radius is gone \citep{vollmer2012, cortese2010}. While decoupled 
molecular clouds have been observed beyond the main gas truncation 
radius in some galaxies \citep[NGC~4402, NGC~4522, 
NGC~4438; ][respectively]{crowl2005, vollmer2008, vollmer2009}, the mass 
in such features is small and they apparently do not survive long. 
Thus, ram pressure stripped spirals in Virgo must be somewhat 
H$_2$-deficient\footnote{\citet{fumagalli2009} has recently reported 
H$_2$-deficient galaxies, although these are anemic galaxies, with low 
surface densities of \ion{H}{i} and H$_2$ out to large radii. Whereas truncated 
galaxies are found almost exclusively in clusters and are clearly ram 
pressure stripped, the origin of the anemic galaxies, which are found 
both inside and outside of clusters, is unclear, although they may be 
starved galaxies.}.

The IC3418 galaxy is highly \ion{H}{i}-deficient, with the deficiency parameter 
def$_{\rm \ion{H}{i}}= \log\ (M_{\rm \ion{H}{i}, orig}/M_{\rm 
\ion{H}{i}, obs})\ga 2.16$ 
\citep{chung2009, gavazzi2005}. The above estimated original amount of 
molecular gas in the galaxy of $\sim 1\times 10^8~M_\odot$ in 
comparison with our CO-based upper limit of $5\times 10^6~M_\odot$ 
indicates that IC3418 could also be poorer in molecular gas by a factor of 
20 (i.e., def$_{\rm H_2}= \log\ (M_{\rm H_2, orig}/M_{\rm H_2, obs}) 
\sim 1.3$). The H$_2$-deficiency is generally small in Virgo spirals 
since ram pressure is not strong enough to strip massive galaxies 
deeply. However, lower mass galaxies with shallower potential wells 
are expected to be more completely stripped in Virgo, and are expected 
to be significantly H$_2$-deficient; IC3418 may be such a galaxy. While 
we cannot presently determine how much of the weakness in CO emission 
is due to H$_2$-deficiency vs. a non-standard CO-to-H$_2$ relation, 
the lack of ongoing star formation in IC3418 strongly suggests the 
galaxy is deficient in molecular gas.

\section{Results -- \chandra\ }\label{resultschandra}
No diffuse X-ray emission is detected from IC3418 and its tail. There 
is also no detection of a shock front ahead of IC3418. The \chandra\ 
data show three point sources in the tail region, which will be 
discussed elsewhere. In this paper, we focus on the limit of the X-ray 
emitting gas in the tail region ($3.5\arcmin \times 1\arcmin$, or 16.8~kpc $\times$ 
4.8~kpc). A 3$\sigma$ upper limit on the X-ray enhancement over the 
Virgo ICM background emission at the tail region is derived from the 
$0.5 - 2$~keV image. A spectral model of the X-ray tail is required to 
convert the count rate limit to the flux limit and the mass limit. The 
X-ray tail is expected to have multiple temperature components. 
However, low statistics of X-ray tails so far have hindered a detailed 
study, so often a single temperature component is assumed \citep[see 
detailed discussions in][]{sun2010}. The known X-ray tails have $T_{\rm 
ICM} / T_{\rm tail}$ ratios of $3 - 8$ \citep{sun2005, sun2010, 
wezgowiec2011}. If we simply take this range for the ratio, the 
expected tail temperature of IC3418 is $0.3 - 0.8$~keV, for a 
surrounding ICM temperature of $\sim 2.5$~keV \citep{shibata2001}. 
In this work, we assume a single APEC model with temperatures of 
$0.3 - 0.7$~keV and abundances of 0.1~solar or 1~solar (see 
Table~\ref{tabX}). The assumed X-ray tail is approximated by a cylinder 
16.8~kpc long with a diameter of 4.8~kpc. The derived upper limits are 
listed in Table~\ref{tabX}. While it is often assumed that the stripped 
ISM is metal rich, a one-$T$ fit always results in very low abundance, 
which is presumably caused by the mixing of multi-$T$ gas in the tail 
\citep[see discussions in][]{sun2010}. 

\begin{table}[!t]
\centering
\caption[]
{Upper limits on the X-ray emission from the IC3418 tail.}
\label{tabX}
\begin{tabular}{ccccc}
\hline\hline
\noalign{\smallskip}
model $T/Z$ & $L_{0.5-2}$ & $L_{\rm bol}$ & $n_e f_{\rm X}^{-1/2}$ & 
$M_{\rm X} f_{\rm X}^{1/2}$\\
(keV/solar) & ($10^{38}$~erg/s) & ($10^{38}$~erg/s) & 
($10^{-3}$~cm$^{-3}$) & ($10^7~M_\odot$)\\
\noalign{\smallskip}
\hline
\noalign{\smallskip}
0.3/0.1 & 3.5 & 13  & 4.3 & 3.7\\
0.3/1.0 & 3.5 & 7.8 & 1.6 & 1.4\\
0.5/0.1 & 3.1 & 7.2 & 3.1 & 2.7\\
0.5/1.0 & 3.0 & 4.5 & 1.2 & 1.1\\
0.7/0.1 & 2.9 & 6.1 & 2.7 & 2.4\\
0.7/1.0 & 2.7 & 4.0 & 1.1 & 0.97\\
\hline
\end{tabular}
\tablefoot{
$L_{0.5-2}$: rest-frame $0.5-2$~keV luminosity. 
$L_{\rm bol}$: X-ray bolometric luminosity. 
$n_e$: electron density. 
$f_{\rm X}$: X-ray volume filling factor. 
$M_{\rm X}$: X-ray mass. 
}
\end{table}

As shown in Table~\ref{tabX}, the derived $0.5 - 2$~keV luminosity 
limit is $\sim 280$ times and $\sim 100$ times lower than the 
luminosities of ESO~137-001 and ESO~137-002, respectively, the spiral 
galaxies in a more distant cluster with prominent ram pressure stripped 
tails with strong X-ray emission \citep{sun2010}. The X-ray mass limit 
is less than $3.7\times 10^7 {\rm f}_{\rm X}^{1/2}~M_\odot$, where 
f$_{\rm X}$ is the filling factor of the X-ray emitting gas in the 
tail. If the X-ray tail in IC3418 is cooler than the 0.3~keV assumed in 
our spectral model, the mass and bolometric luminosity limits would 
become weaker. 

In numerical simulations the amount of X-ray emitting gas in the tail 
was found to depend strongly on the surrounding ICM pressure 
\citep{tonnesen2011}. The thermal pressure of the ICM at the location 
of IC3418 in Virgo is $P_{\rm ICM}= n_{\rm ICM} T\sim 2.8\times 
10^{-12}$~dyne\,cm$^{-2}$, where $n_{\rm ICM}= 7.1\times 
10^{-4}$~cm$^{-3}$ is the local ICM density (see Sect.~\ref{RPS} for 
details), and $k_B T= 2.5$~keV is the local ICM temperature 
\citep{shibata2001}. The simulation run of \citet{tonnesen2010} 
with a close ICM thermal pressure value of $1.76\times 
10^{-12}$~dyne\,cm$^{-2}$ suggests that some soft ($0.5-2$~keV) X-ray 
emission could be observable in the tail of IC3418 for our sensitive 
\chandra\ surface brightness limit of $\sim 4\times 
10^{-7}$~erg\,s$^{-1}$\,cm$^{-2}$. This follows from Fig.~2 (right 
panel) in \citet{tonnesen2011}, assuming the same column density of 
about $10^{20}$~cm$^{-2}$ for the hot gas in the tail of IC3418 as is 
measured in the simulations. However, the gas mass and gas density in 
the tail are expected to strongly evolve over time, and 
\citet{tonnesen2010} did not study the evolutionary stage after the 
galaxy is fully stripped, like in IC3418. 

Once the main body of the galaxy has been (nearly) completely stripped, 
the source of new tail gas is depleted. In one of the simulations of 
\citet{kapferer2009}, who use a different numerical hydrodynamics 
scheme than does Tonnesen (smoothed particle hydrodynamics, SPH, vs. 
adaptive mesh refinement, AMR), a model massive galaxy is 
completely stripped by a strong ram pressure. Some 400~Myr after the 
galaxy was stripped, there is still a very long bright X-ray tail that 
is spatially coincident with a star-forming tail. However, IC3418 is a 
small galaxy, originally with a much smaller amount of gas. Thus, after 
the main body has been nearly completely stripped, the gas density of 
the tail is expected to drop. It is possible that the X-ray tail lags 
farther behind the SF tail which is the case in ESO~137-001, for 
example, where the brightest X-ray region is 35~kpc from the galaxy. 
However, the Chandra FOV covers another $\sim 2.2\arcmin$ after the end of 
the SF tail, with no sign of the X-ray enhancement. The X-ray surface 
brightness is about the same, so the flux limit would be $\sim 28\%$ 
higher if a region of $5.7\arcmin\times 1\arcmin$ is considered (instead of 
$3.5\arcmin\times 1\arcmin$). 

Thus, one reason for the weak X-ray (as well as H$\alpha$ and 
\ion{H}{i}) emission in the tail of IC3418 could be that it is in an 
advanced evolutionary stage when the gas density of the tail is 
expected to drop. In the Virgo cluster, however, almost no X-ray tails 
are observed in other ram pressure stripped galaxies with \ion{H}{i} or 
H$\alpha$ tails. The possible reasons for our non-detection thus 
further include low pressure of the surrounding ICM (for comparison, 
the ICM thermal pressure at the position of ESO~137-001 in the A3627 
cluster is about $1.8\times 10^{-11}$~dyne\,cm$^{-2}$, i.e., about six 
times higher than in Virgo at the position of IC3418) or a lower tail 
temperature than we assumed in the spectral model, which would also 
mean fast cooling (assuming a not-very-small abundance) and thus a 
short lifetime. 

\section{Ram pressure stripping}\label{RPS}
Ram pressure stripping is very likely the principal cause of the strong 
gas deficiency of IC3418 and of its peculiar morphology 
\citep{hester2010, fumagalli2011}. Acceleration due to ram pressure 
basically depends on the column density of the gas parcels and the 
gravitational potential of the galaxy. The shallow potential of the 
dwarf makes its interstellar matter more susceptible to ram pressure 
and so material with higher column density can be affected. Compared to 
other ram pressure stripped galaxies in the Virgo cluster, IC3418 is 
also projected closer to the cluster center where the ICM density is 
higher. Although we don't know the true de-projected distance, IC3418 
is probably affected by a stronger ram pressure than other Virgo 
galaxies. 

In this section we use semi-analytic calculations to model the 
effects of a time-varying ram pressure on both atomic and molecular ISM 
components in a simple manner by taking into account their column 
density as a critical parameter. In a static potential of a 
two-component model of the galaxy consisting of the disk and halo 
(the contribution of a bulge is neglected), we follow dynamics of ISM 
parcels at different disk radii under the influence of an external 
time-varying face-on force. Every parcel is assigned a column density. 
For every parcel we calculate the equation of motion in the vertical 
direction that takes into account local ram pressure and the restoring 
force from the galaxy. The local ram pressure may be expressed as 
$d\upsilon/dt= - \rho_{\rm ICM}\, |\upsilon - \upsilon_0|^2/ 
\Sigma_{\rm ISM}$, where $\Sigma_{\rm ISM}$ is the mean column density 
of an ISM parcel and $(\upsilon - \upsilon_0)$ is the vertical 
component of its relative velocity with respect to the surrounding hot 
gas. The galaxy is modeled with a Miyamoto-Nagai disk and a Plummer 
halo \citep[e.g.,][]{binney2008} with the parameters specified in 
Table~\ref{TabModel}.

\begin{table}[t]
\caption[]
{Parameters of galaxy and cluster models used for semi-analytic 
calculations of ram pressure stripping.}
\label{TabModel}
\begin{tabular}{lll}
\hline
\hline
\noalign{\smallskip}
galaxy:  & $M_\mathrm{d}= 5\times 10^8~M_\odot$ & $a_\mathrm{d}= 1.5$~kpc \\
         & $M_\mathrm{h}= 7\times 10^9~M_\odot$ & $a_\mathrm{h}= 3$~kpc  \\
\noalign{\smallskip}
\hline
\noalign{\smallskip}
cluster: & $M_\mathrm{vir}=1.4\times10^{14}~M_\odot$ & $r_\mathrm{s}=320$~kpc\\
	 & $\delta_\mathrm{th}=340$ & $\Omega_\mathrm{m}=0.3$\\
	 & $\rho_{\rm ICM,1}(0)= 4.4\times 10^{-22}$~kg\,m$^{-3}$ & $r_{c,1}=1.2$~kpc\\
	 & $\rho_{\rm ICM,2}(0)= 1.4\times 10^{-23}$~kg\,m$^{-3}$ & $r_{c,2}=23.7$~kpc\\
	 & $\beta_1=0.42$ & $\beta_2=0.52$\\
\noalign{\smallskip}
\hline
\end{tabular}
\end{table}

At the projected distance of IC3418 from M87 of about $1\degr$, the 
density of the ICM is $\sim 7.1\times 10^{-4}$~cm$^{-3}$, assuming a 
spherically symmetric and smooth distribution fitted with a double 
$\beta$-profile \citep{matsushita2002}. The parameters used are given 
in Table~\ref{TabModel}. We get an upper limit on the local ram 
pressure of about 1820~cm$^{-3}$(km\,s$^{-1}$)$^2 = 1.8\times 
10^{-11}$~dyne\,cm$^{-2}$, assuming a deprojected orbital velocity of 
1600~km\,s$^{-1}$ for IC3418. The time evolution of the ram pressure 
exerted on the disk corresponds to a profile along an orbit of IC3418 
that is consistent with its observed plane of the sky position with 
respect to the Virgo cluster center and line-of-sight velocity. The 
projected tail direction was assumed to indicate the current direction 
of motion in the plane of the sky. We have studied orbits with 
peri- to apocenter ratios from about 1:5 to 1:20 that are characteristic 
for radial orbits in galaxy clusters \citep[][]{ghigna1998, 
boselli2006, vollmer2009}. As a fiducial orbit we have chosen an orbit 
that brings IC3418 currently almost to pericenter, about 275~kpc from 
the cluster center, with the plane of the sky velocity components 
$\upsilon_x = \upsilon_y = 820$~km\,s$^{-1}$. The total 3D velocity is 
thus $\sim 1600$~km\,s$^{-1}$. The gravitational potential of the Virgo 
cluster was approximated by the spherically symmetric potential of the 
NFW dark matter halo \citep{navarro1996} with parameters specified in 
Table~\ref{TabModel}. 

\begin{figure}[t]
\centering
\includegraphics[height=0.45\textwidth,angle=270]{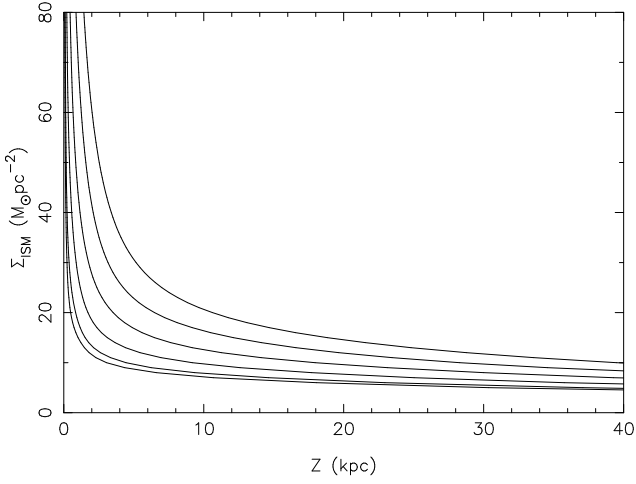}
\caption{
Pre-peak stripping: vertical distances that ISM elements can reach with 
different column densities from radii $0-5$~kpc (with 1~kpc step; 
\textit{from left to right}) in our model 100~Myr before the peak ram 
pressure. 
}\label{FigSigISM-prepeak}
\end{figure}

In Figs.~\ref{FigSigISM-prepeak} and \ref{FigSigISM-postpeak}, the 
vertical distance behind the disk to which parcels with different 
column densities occurring at different radii (up to 5~kpc; the $R_{25}$ 
major axis radius of IC3418 is $\sim 3.6$~kpc; see Table~\ref{ic3418}) 
can get in our model are shown for two time moments along the ram 
pressure event: 100~Myr before the peak and 100~Myr after the peak. In 
the pre-peak situation, only parcels with $\Sigma_{\rm ISM}< 
15~M_\odot$\,pc$^{-2}$ can be shifted (not necessarily stripped) to at 
least 20~kpc, i.e., to distances comparable to the projected length of 
the tail of IC3418. From the nucleus of the galaxy, parcels with 
$\Sigma_{\rm ISM}\approx 5-10~M_\odot$\,pc$^{-2}$ can get to such 
distances. In the post-peak situation, more material is accelerated to 
larger distances -- parcels from outer disk radii with $\Sigma_{\rm 
ISM}$ up to about $35~M_\odot$\,pc$^{-2}$, and from inner radii with 
$\Sigma_{\rm ISM}< 15~M_\odot$\,pc$^{-2}$, can get to $z\sim 20$~kpc.

Figures~\ref{FigSigISM-prepeak} and \ref{FigSigISM-postpeak} thus 
suggest that in our simulation low-density (\ion{H}{i}-like) gas is 
stripped from the galaxy quite easily and can get to large distances; 
in the post-peak situation, all the gas with $\Sigma_{\rm ISM}\la 
5~M_\odot$\,pc$^{-2}$ gets to $z> 100$~kpc. Between the inner and outer 
radii, there is a large difference in the column density of parcels 
that can be moved to a certain distance. Denser gas parcels are thus 
difficult to strip substantially. 

Figures~\ref{FigSigISM-prepeakVz} and \ref{FigSigISM-postpeakVz} depict 
the vertical velocities relative to the galaxy to which ISM elements can 
be accelerated depending on their column density, for both the pre-peak 
and post-peak stages. Only low-density stripped parcels can reach high 
velocities in the pre-peak phase and the situation does not change much 
in the post-peak phase where elements with column densities 
$<15~M_\odot$\,pc$^{-2}$ can occur at about 400~km\,s$^{-1}$ relative to 
the galaxy. This excludes the possibility that some weak features noted 
in the spectra of the K4 and K6 tail positions in Sect.~\ref{iramtail} 
may correspond to a real CO emission. 

\begin{figure}[t]
\centering
\includegraphics[height=0.45\textwidth,angle=270]{./Fig8.ps}
\caption{
Post-peak stripping. Same as in Fig.~\ref{FigSigISM-prepeak}, but 
100~Myr after the peak ram pressure.
}\label{FigSigISM-postpeak}
\end{figure}
\begin{figure}[t]
\centering
\includegraphics[height=0.45\textwidth,angle=270]{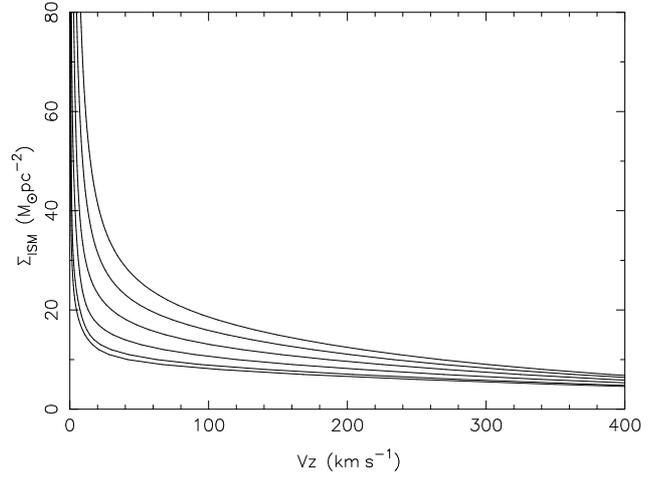}
\caption{
Pre-peak stripping: vertical velocities to which ISM elements can be 
accelerated with different column densities from radii $0-5$~kpc (with 
1~kpc step; \textit{from left to right}) in our model 100~Myr before the 
peak ram pressure. 
}\label{FigSigISM-prepeakVz}
\end{figure}
\begin{figure}[t]
\centering
\includegraphics[height=0.45\textwidth,angle=270]{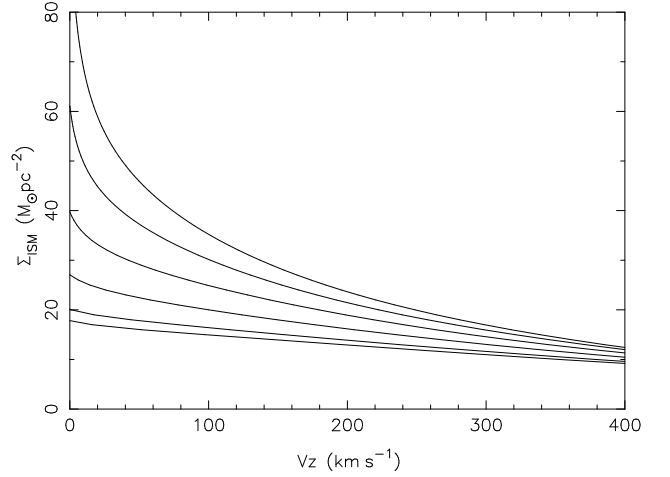}
\caption{
Post-peak stripping. Same as in Fig.~\ref{FigSigISM-prepeakVz}, but 
100~Myr after the peak ram pressure.
}\label{FigSigISM-postpeakVz}
\end{figure}

\begin{figure}[t]
\centering
\includegraphics[height=0.47\textwidth,angle=270]{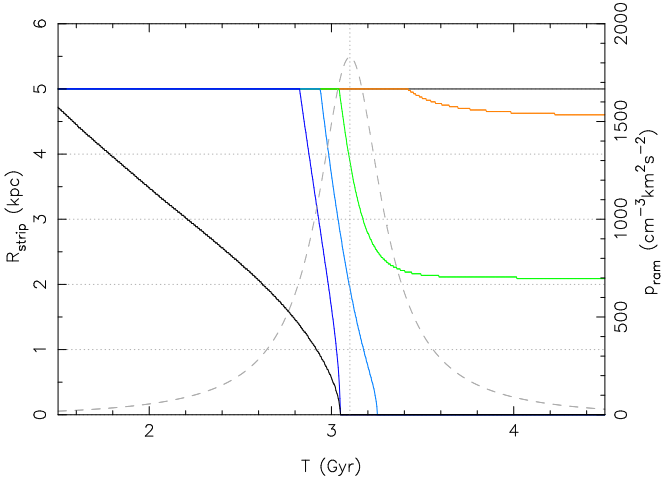}
\caption{
Time evolution of stripping radii of ISM parcels with different column 
densities: \ion{H}{i}-like (radius-dependent Miyamoto-Nagai profile with 
$\Sigma_0= 10~M_\odot$\,pc$^{-2}$), and constant $\Sigma_{\rm ISM}=10$, 
15, 25, and 50~$M_\odot$\,pc$^{-2}$ (\textit{from left to right}) 
in our model. Ram pressure time profile along our model orbit is shown 
with the dashed line. 
}\label{Figrstrip-time}
\end{figure}

In Fig.~\ref{Figrstrip-time} we show the time evolution of the 
stripping radii of ISM parcels with different column densities. The 
stripping radius of a disk gas component with some $\Sigma_{\rm ISM}$ 
corrensponds to a radius outside which it is released from the 
galaxy's potential ($E_{\rm tot}= E_{\rm kin}+ E_{\rm pot} >0$). 
Figure~\ref{Figrstrip-time} provides a timescale for the stripping along 
our fiducial orbit: \ion{H}{i}-like parcels (modeled with a 
Miyamoto-Nagai profile with $\Sigma_0= 10~M_\odot$\,pc$^{-2}$) start to 
be released from the outer disk by a weak ram pressure at cluster 
outskirts. It takes about 1~Gyr before they are removed completely 
from $R=3$~kpc to the center of the galaxy. Denser gas parcels, with 
$\Sigma_{\rm ISM}$ up to $\sim 20~M_\odot$\,pc$^{-2}$, can also be 
stripped throughout the disk. This happens on shorter timescales of 
$100-200$~Myr from $R=3$~kpc to 0~kpc as higher ram pressure arising 
only closer to the pericenter is required. For yet higher values 
of the column density the stripping radius starts to increase and 
subsequently approaches the outer disk radius. This happens at 
$\Sigma_{\rm ISM}\sim 50~M_\odot$\,pc$^{-2}$ which suggests that parcels 
with densities at this or a higher level can never be stripped from the 
galaxy by ram pressure alone. 

Our simulation results are consistent with the results of spectral 
modeling of observations which suggested that star formation in the 
main body of IC3418 stopped about $200-300$~Myr ago, on a timescale of 
less than $\sim 70$~Myr from the outside in (Kenney et al., in prep.). Some 
of the denser gas in Fig.~\ref{Figrstrip-time} is released only after 
maximum ram pressure, but this is part of material that had already 
been pushed out of the disk plane to the halo region, and will eventually 
fall back to the disk \citep[see][for details]{jachym2009}. Although 
our calculations depend on the adopted values of the free orbital 
parameters not constrained by observations and on the total mass of the 
galaxy, the results are consistent with what is expected from other 
simulations and existing observations of galaxies in clusters of the 
size of Virgo \citep{kenneyyoung1989, boselli2002, fumagalli2008}: it 
is not possible to directly strip the dense (molecular) clouds, so they 
must have been removed from IC3418 in some other way that we will 
discuss later. 

\section{Discussion}\label{discussion}

\subsection{Central bulk of molecular gas}\label{centralbulk}
If our marginal detection of CO emission in the B1 pointing is real, it 
corresponds to gas concentrated in the nuclear region. The possible 
survival of dense gas in the galaxy core (central few arcseconds = few 
hundred pc) could either be because it is the center of the galaxy, the 
deepest part of the potential well from which it is most difficult to 
remove gas, or because the surface density of molecular gas is higher 
than the rest of the ISM, or both. While some molecular gas in the 
nucleus has apparently survived the stripping, there is no ongoing star 
formation in this gas. This might be due to a gas density that is below 
a threshold needed for star formation. The detected gas survived $\sim 
200$~Myr after the cessation of star formation (Kenney et al., in 
prep.) that is presumably due to removal of most of the ISM. 

The survival of molecular gas in the very central region of an 
otherwise gas-stripped dwarf irregular galaxy could be relevant for the 
formation of nucleated dwarf elliptical galaxies. Many cluster dEs have 
distinct nuclear star clusters, and it is not understood why some dEs 
have them and others do not. A surviving central gas reservoir may make 
it possible to form a nuclear star cluster even after the galaxy is 
mostly gas-stripped. Also, as shown in numerical simulations, 
high-density clouds may lose angular momentum through interaction with 
the ICM and drift toward the center of the galaxy \citep{tonnesen2009}. 

The heliocentric central velocity of the marginally detected central CO 
line is about 100~km\,s$^{-1}$ (see Sect.~\ref{irammain}). From Keck 
spectra, Kenney et al. (in prep.) measured a value of $170\pm 
10$~km\,s$^{-1}$ for the central velocity of the stars in the galaxy. 
The galaxy's rotation is presumably too small to account for such a 
velocity difference. Moreover, acceleration due to ram pressure would 
push the gas toward a higher radial velocity. On the other hand, if 
dense gas clouds can fall back into the galaxy after they condense in 
the tail and then become decoupled from ram pressure, they would have 
blue-shifted radial velocities relative to the central stellar disk. 
This behavior was noted in ram pressure stripping simulations of 
\citet{kapferer2009} and could work in IC3418. Future more sensitive CO 
observations could resolve the origin of the observed mismatch between 
the velocities of the stellar and molecular components. 

\subsection{Molecular gas and star formation in the tail}\label{tail}
To assess the current molecular content of the tail of IC3418, we 
inquire how much molecular gas is needed for the amount of star 
formation in the tail observed. Using the definition of GALEX AB 
magnitudes\footnote{
$m_{\rm AB}= -2.5 \log({f_\nu} [{\rm erg\,cm^{-2}\,s^{-1}\,Hz^{-1}}]) - 
48.6$}, the total FUV luminosity of the tail is about $3.6\times 
10^{25}$~erg\,s$^{-1}$\,Hz$^{-1}$ \citep{fumagalli2011}. Making the 
assumption of a continuous star formation over the last $\sim 100$~Myr, 
Eq.~\ref{kennicutt} predicts a SFR of $\sim 3.4\times 
10^{-3}~M_\odot$\,yr$^{-1}$. The total H$\alpha$ luminosity of the 
knots in the outer half of the tail of about $2.4\times 
10^{38}$~erg\,s$^{-1}$ \citep{fumagalli2011} corresponds to a SFR of 
$\sim 1.3\times 10^{-3}~M_\odot$\,yr$^{-1}$ over the last $\sim 10$~Myr 
\citep[][]{kennicutt2012}\footnote{
$\log{\rm SFR}\ (M_\odot\,{\rm yr}^{-1})= \log{L_{\rm H\alpha}}\ ({\rm 
erg\,s^{-1}}) - 41.27$}. The difference in SFR estimates is not large 
and may be mainly because H$\alpha$ emission is restricted 
to the outer part of the tail. Adopting a SFR in the tail of $\sim 
3\times 10^{-3}~M_\odot $\,yr$^{-1}$, and assuming a typical star 
formation timescale (or star formation efficiency ${\rm SFE(H_2)= 
\Sigma_{SFR}/ \Sigma_{H_2}}$) of $\sim 2$~Gyr 
\citep[e.g.,][]{bigiel2008, leroy2013} we can deduce the mass of 
molecular gas expected to be occurring in the tail to be $6\times 
10^6~M_\odot$. Our observed ($3\sigma$) upper limit on CO amount in the 
tail (see Table~\ref{limits}) is thus consistent with that predicted for 
a normal gas depletion timescale and normal X-factor. We would need 
more sensitive CO observations to show that either of these was 
non-standard. 

\begin{table*}[!t]
\centering
\caption[]{
Star formation rates in the observed tail regions and corresponding 
lower limits on local CO-to-H$_2$ conversion factor. FUV and H$\alpha$ 
luminosities come from \citet{fumagalli2011}. FUV- and H$\alpha$-based 
SFRs were calculated using the formula of \citet{kennicutt2012}. For the 
X-factor estimates, a 2~Gyr H$_2$ depletion time was assumed.}
\label{tailSFR}
\begin{tabular}{lcccccc}
\hline\hline
\noalign{\smallskip}
Source & $L_{\rm FUV}$ & $L_{\rm H\alpha}$ & SFR$_{\rm FUV}$ & 
SFR$_{\rm H\alpha}$ & $L_{\rm CO}$ & $\alpha_{\rm CO}$\\
       & ($10^{39}$~erg\,s$^{-1}$) & ($10^{37}$~erg\,s$^{-1}$) & 
($10^{-4}~M_\odot$\,yr$^{-1}$) & ($10^{-4}~M_\odot$\,yr$^{-1}$) & 
($10^5$~K\,km\,s$^{-1}$\,pc$^2$) & 
($M_\odot$\,pc$^{-2}$\,(K\,km\,s$^{-1}$)$^{-1}$)\\
\noalign{\smallskip}
\hline
\noalign{\smallskip}
disk & 139.3& -   & 62.2 & -   & $<11$  & - \\
K4   & 7.7  & 8.0 & 3.5  & 4.3 & $<2.6$ & $>3.3$ \\
K6   & 8.6  & 5.2 & 3.8  & 2.8 & $<2.3$ & $>2.4$ \\
K8   & 4.6  & 2.7 & 2.1  & 1.5 & $<2.1$ & $>1.4$ \\
\hline
\end{tabular}
\end{table*}

Table~\ref{tailSFR} summarizes FUV and H$\alpha$ luminosities for the 
main body and the three tail regions that we observed in CO, and 
compares the corresponding SFRs. It shows that SFR estimates from the 
two tracers are in the knots similar, suggesting that indeed the 
difference in these quantities for the whole tail is because star 
formation stopped in parts of the tail. Applying our upper limit CO 
luminosities in the observed regions, we can estimate from the 
H$\alpha$ SFRs lower limit values on the CO-to-H$_2$ conversion factor 
$\alpha= \tau_{\rm dep}\times {\rm SFR}/L_{\rm CO}$, assuming a 
constant depletion time $\tau_{\rm dep}= 2$~Gyr. The resulting 
limits are consistent with the Galactic value $\alpha_{\rm CO}\sim 
4.4~M_\odot $\,pc$^{-2}$\,(K\,km\,s$^{-1}$)$^{-1}$. 
\citet{vollmer2012} and \citet{boissier2012} determined that the star 
formation efficiency in the gas stripped tails of Virgo galaxies with 
respect to the available amount of atomic gas was at least 10 times 
lower than within the galaxies. However, \citet{vollmer2012} found normal 
values of the SFE with respect to molecular content in the extraplanar 
regions of the galaxies. 

\subsection{Where is the gas?}\label{whereis}
Although dwarf galaxies are gas-rich systems, in IC3418 practically no 
gas, neither atomic \ion{H}{i} nor molecular H$_2$, is observed. No 
\ion{H}{i} is detected in the body and the tail, with fairly sensitive 
upper limits, $M_{\rm \ion{H}{i}}< 8\times 10^6~M_\odot$ \citep{chung2009}. This is 
$\sim 1\%$ of the \ion{H}{i} mass expected in a typical late type dwarf galaxy 
with the optical size or $H$-band luminosity of IC3418. We have 
possibly found only a small amount of molecular gas $M_{\rm H_2} \simeq 
10^6~M_\odot$ in the central $11\arcsec$ of the main body, and only upper 
limits ($M_{\rm H_2}< 10^7~M_\odot$ assuming standard CO-to-H$_2$ 
conversion) outside the central region and in the tail. While our upper 
limits are not sensitive enough to clearly demonstrate a lack of 
molecular gas in the tail (see Sect.~\ref{tail}), there is very likely a 
large amount of molecular gas missing in the main body. Although this is 
consistent with the observed absence of any ongoing star formation in 
the body of the galaxy, the question arises of what happened to the 
molecular gas that once occurred in IC3418. 

Calculations in the previous section have shown that low-density 
(atomic) gas can easily be completely stripped by ram pressure from a 
model of IC3418, which is consistent with the sensitive non-detection 
of any \ion{H}{i} emission. Higher-density gas with $\Sigma_{\rm ISM}$ 
up to $\sim 50~M_\odot$\,pc$^{-2}$ then can be accelerated from outer 
radii only, while the densest (molecular) gas stays 
bound to the galaxy. However, stripping by dynamical pressure of the 
ICM is probably enhanced by other processes such as ablation of dense 
clouds disrupted by hydrodynamic instabilities 
\citep[e.g.,][]{quilis2000, kapferer2009, tonnesen2009}, or 
shock-heating and dissociation of molecular clouds due to supersonic 
motion of IC3418 through the Virgo ICM \citep[e.g.,][]{guillard2012}. 
Moreover, the structure of the ISM in dwarf irregulars can be 
significantly influenced by stellar feedback and often contains up to 
$\sim$kpc-scale shells and holes seen in \ion{H}{i} \citep[see, 
e.g.,][and references therein]{zhang2012}, which could facilitate 
stripping of the ISM as was shown in numerical simulations of 
\citet{quilis2000} with holey disks. Thanks to a shallow potential well 
and low stellar surface densities, the ISM in dwarfs can also have 
lower average surface densities than in 
large spirals. Besides these indirect processes, a very important 
effect may also come from the rather limited lifetimes of dense clouds 
($<100$~Myr) that are comparable to the typical timescales of ram 
pressure stripping. The molecular gas reservoir could not then be 
replenished by further condensation and molecularization of \ion{H}{i} 
since that was removed from the galaxy. All these processes, when 
combined together, may have resulted in effective stripping of 
molecular ISM in IC3418. As a result, IC3418 may currently be virtually 
deficient of almost all gas except possibly in its very center. 

\subsection{Gas phases in the tail}

\begin{table}[!t]
\centering
\caption[]{
Upper limit ($3\sigma$) masses (in $M_\odot$) of gas components issued 
from different observations of the tail. 
}
\label{tabUpper}
\begin{tabular}{lcccc}
\hline\hline
\noalign{\smallskip}
 & H$_2$ & \ion{H}{i} & H$\alpha$ & X-rays\\
\noalign{\smallskip}
\hline
\noalign{\smallskip}
$M_{\rm tail}$ & $<1\times 10^7$ & $<4\times 10^8$ & $< 4\times 
10^8\ {\rm f}_{\rm H\alpha}^{1/2}$ & $<4\times 10^7\ {\rm f}_{\rm 
X}^{1/2}$\\
ref. & this work & Ch09 & this work\tablefootmark{a} & this work\\
\hline
\end{tabular}
\tablefoot{Ch09 = \citet{chung2009}; \tablefoottext{a} Based on 
H$\alpha$ surface limit from Kenney et al. (in prep.) 
}
\end{table}

Table~\ref{tabUpper} summarizes the upper limits on the amount of different 
gas phases observed in the tail. Once stripped, the gas has probably 
been heated to temperatures between that of \ion{H}{i} and the ICM, as cooler 
gas in the tail mixes with hot ICM gas, or more dense gas has been 
condensing into star-forming clouds. The absence of bright X-ray 
emission (see Sect.~\ref{resultschandra}) is likely because of a low gas 
content in the tail associated with its advanced evolutionary state 
and/or because of rather low thermal pressure of the ICM at the position 
of IC3418 in the Virgo cluster. Compression of stripped gas by the 
surrounding ICM determines the temperature and density distribution of 
gas in the tail. If present, the hot gas component in the tail is thus 
probably too diffuse to be observable in X-rays. The actual upper limit 
mass is close to $\sim 4\times 10^7~M_\odot$ since the filling factor 
of the X-ray emitting gas is not expected to be much smaller than 1. 
Associated with the \ion{H}{ii} regions, there is clearly ionized gas in the 
tail, but it is difficult to reliably estimate its mass without knowing 
the gas density; however the ionized gas mass in \ion{H}{ii} regions is 
generally small. While no diffuse H$\alpha$ emission has been detected, 
there could be significant mass associated with undetected diffuse 
$10^4$~K gas. We derive the upper mass limit on H$\alpha$ emitting 
diffuse gas from the $2\sigma$ limit of the H$\alpha$ surface 
brightness of $1.36\times 
10^{-17}$~erg\,s$^{-1}$\,cm$^{-2}$\,arcsec$^{-2}$ observed with 
WIYN~3.5m telescope (Kenney et al., in prep.). This is $\sim 2.3$ times 
more sensitive than the H$\alpha$ image presented by 
\citet{fumagalli2011}. If the H$\alpha$ emission is considered to 
originate from the same cylindrical volume of the X-ray gas, assuming 
an electron temperature of $10^4$~K and case B of nebular theory, the 
mass limit of the H$\alpha$-emitting gas is $2.8\times 10^8~{\rm 
f}_{\rm H\alpha}^{1/2}~M_\odot$, where f$_{\rm H\alpha}$ is the filling 
factor of the H$\alpha$ gas. Numerical simulations of ram pressure 
stripping indicated that bright H$\alpha$ emission is produced at the 
edges of dense neutral clouds \citep{tonnesen2011}, so the volume 
filling factor should be low. From numerical simulations of 
\citet{tonnesen2011} with the value of ICM thermal pressure consistent 
with that for ESO~137-001 (their Fig.~2, left panel), it follows that 
the typical local density of the H$\alpha$ gas in the ESO~137-001 tail 
is $\sim 1$~cm$^{-3}$. Since the rms electron density observed in the 
tail of ESO~137-001 is $\sim 0.045$~cm$^{-3}$ \citep{sun2007}, the 
filling factor, defined as the ratio of the average density and local density, 
is estimated to be about $5\times 10^{-2}$. Thus, a reasonable upper 
limit on the amount of H$\alpha$ emitting gas in the tail of IC3418 
would be about $6\times 10^7~M_\odot$. Considering the atomic component 
of the gas in the tail, the mass upper limit is $\sim 4\times 
10^8~M_\odot$ following from the upper limit of $8\times 10^6~M_\odot$ 
per beam \citep[FWHM $\sim 16\arcsec$,][]{chung2009}, and assuming the same 
tail dimensions used above for the X-ray mass estimate (see 
Sect.~\ref{resultschandra}). The actual amount of the \ion{H}{i} in the tail 
may be much lower than the limit since denser clouds have probably 
condensed into stars while the diffuse gas had been stripped to large 
distances (very likely exceeding the observed length of the tail), as 
follows from the calculations in Sect.~\ref{RPS}. Numerical simulations of 
\citet{tonnesen2011} suggest that in the rather low thermal pressure 
environment surrounding IC3418, the mass fractions of cool ($300~{\rm K} 
< T < 10^4$~K) and hot ($7\times 10^5~{\rm K} < T < 4\times 10^7$~K) 
components in the stripped gas might be both of about 30\% of the total 
stripped mass without much evolution as the tail ages from 100~Myr to 250~Myr. 

\subsection{Origin of star formation in the tail}
Galaxies with star forming tails are rare in the Virgo cluster. 
Extraplanar H$\alpha$ emission was found in only a few ram pressure 
stripped galaxies \citep[NGC~4522, NGC~4402, NGC~4330, 
NGC~4438,][respectively]{kenney1999, cortese2003, abramson2011, 
boselli2005} where it typically occurs only fairly close to the disk. 
Those are galaxies with typical (stellar) mass at least a factor 
of 10 higher than that of IC3418. In IC3418 the situation is different 
-- a large number of young stars occur at quite large distances (up to 
17~kpc in projection) from the disk distributed in a straight, narrow 
tail. \citet{chung2009} suggested that recent star formation in the 
tail occurred from compression of the stripped \ion{H}{i} gas. Similarly, 
\citet{hester2010} speculated that star formation in the wake was 
driven by turbulence that enhances density contrasts and thus aids gas 
cooling and condensation. A similar example of star formation in situ 
in the stripped gas is known in the Virgo dwarf galaxy VCC1249 
\citep{arrigoni2012} which is, however, stripped by the ISM of an 
elliptical galaxy (M49), whereas IC3418 is stripped by the ICM in the 
Virgo cluster.

Hydrodynamic AMR simulations of \citet{tonnesen2012} showed that star 
formation may occur in the ram pressure stripped tails when radiative 
cooling can overcome the heating from the surrounding ICM and thus is 
able to cool down the stripped ISM. Our simplified calculations in 
Sect.~\ref{RPS} have suggested that although molecular gas cannot be 
directly stripped by ram pressure from IC3418, some higher density \ion{H}{i} 
than is usual in larger spirals in Virgo ($\Sigma_{\rm ISM}< 
25~M_\odot$\,pc$^{-2}$ within the inner 2~kpc disk radius, and $< 
50~M_\odot$\,pc$^{-2}$ in the outer disk radii) can be stripped thanks 
to a shallow potential of the galaxy and its proximity to the cluster 
center. It is then possible that these higher-density clumps were easier 
to cool and compress by ram pressure together with the thermal pressure 
of the ICM, supporting their condensation. Moreover, because of the 
supersonic motion of IC3418 through the ICM, the thermal pressure of the 
post-shock environment surrounding the stripped ISM within its Mach 
cone is probably higher than the pre-shock thermal pressure, which could 
also enhance condensation of the stripped material. In simulations, the 
stripped ISM fragments strongly behind the bow shock as a result of turbulence 
and Rayleigh-Taylor instability \citep[e.g.,][]{roediger2006}. As 
shown, e.g., by \citet{zhang2012}, the inner disks in nearby dwarf irregulars 
have proportionally more cool gas than the outer disks. 
\citet{leroy2008} indeed suggested that H$_2$ dominates ISM of nearby 
dwarfs in the inner parts of their disks ($<0.25 r_{25}$) and not at 
the outer radii. Then ram pressure stripped cool gas and denser clumps 
originate preferentially from the inner disk regions. The global 
head-tail morphology of IC3418 with a narrow (and linear) tail 
extending behind a larger main body of the galaxy, would then be 
consistent with the proposed scenario.

\section{Conclusion}\label{conclussion}
We searched for cold molecular gas, as well as hot X-ray emitting 
gas, in IC3418, a dwarf galaxy that is currently being transformed by 
ram pressure stripping from an irregular to an early-type galaxy. It contains 
a prominent tail of young stars, a feature that is rare among other ram 
pressure stripped Virgo galaxies. Using the IRAM 30m antenna, we 
searched for CO emission in five deep integrations in the main body of 
the galaxy and in its gas-stripped tail. New deep \chandra\ 
observations covered the whole system. The main results are: 

\begin{enumerate}
\item A possible $^{12}$CO(2-1) $5\sigma$ marginal detection suggested 
that about $1.2\times 10^6~M_\odot$ of molecular gas (assuming a standard 
Galactic CO-to-H$_2$ conversion factor) is present in the central 
few arcseconds of IC3418. Although CO(1-0) emission was not formally 
detected in the center of the galaxy, there is a $2\sigma$ feature with 
the same velocity range as the CO(2-1) line, which is consistent with a 
weak emission diluted by the CO(1-0) large beam. An upper limit on 
molecular gas content in the whole main body of the galaxy is estimated 
to $<5\times 10^6~M_\odot$, assuming the standard Galactic CO-to-H$_2$ 
conversion factor. A similar limit is obtained from the existing upper 
limits on FIR emission from dust. This is a factor of $\sim 20$ less 
than the original molecular gas content of the galaxy, as estimated 
from the UV-based star formation rate before quenching, and the typical 
gas depletion timescale in spirals of 2~Gyr. While some of this 
difference may arise from a non-standard CO-to-H$_2$ conversion that 
could be $\sim 5$ times larger than Galactic, we think much of it is 
due to a true deficiency of molecular gas, given the lack of star 
formation in the main body. 

\item Although the presence of \ion{H}{ii} regions in the tail suggests that 
there must be some dense (presumably molecular) gas in the tail, we 
have found only ($3\sigma$) upper limits of $<10^6~M_\odot$ in the 
three observed points covering the outer tail regions. The estimated 
upper limit on the molecular gas content of the whole tail is $\sim 
1\times 10^7~M_\odot$. This is an amount that is similar to the 
estimates based on the observed SFR averaged over the tail, assuming a 
normal gas depletion timescale of 2~Gyr. Thus with our current 
sensitivity we cannot claim that the relationship between the dense gas 
content and the star formation rate in the tail is unusual. 

\item Very sensitive upper limits on the $0.2-5$~keV luminosity of the 
gas-stripped tail has been obtained with \chandra. The corresponding 
X-ray mass limit of the tail is $< 3.7\times 10^7 {\rm 
f_X^{1/2}}~M_\odot$ (where $\rm f_X$ is the filling factor of the X-ray 
emitting gas) which is a factor of $\sim 280$ less than in the tail 
of the Norma cluster ram pressure stripped galaxy ESO~137-001. The 
absence of a bright X-ray tail may be mainly due to a low gas content 
in the tail, associated with its advanced evolutionary state in which 
gas from the main body of the galaxy is no longer supplied to the tail. 
A rather low thermal pressure of the surrounding ICM in the Virgo 
cluster could also account for our non-detection. Existing AMR 
hydrodynamic numerical simulations would suggest that at least some 
soft X-ray emission should be observable in the tail of IC3418; 
however, these models don't study the evolutionary state after the 
galaxy is fully stripped. 

\item Our semi-analytic calculations have indicated that ISM parcels 
with rather high column densities up to $\sim 25- 
50~M_\odot$\,pc$^{-2}$ (depending on the galactic radius) can be 
accelerated from the galaxy. However, molecular clouds with densities 
$>100~M_\odot$\,pc$^{-2}$ cannot be directly stripped from the inner 
part of IC3418 by the ram pressure in Virgo, even though this dwarf 
galaxy has a shallow galactic potential, and it may get close to the 
cluster center. This explains why some molecular gas could survive in 
the galaxy center. Moreover, the apparent loss of most of the molecular 
gas in the galaxy must be due to rather short lifetimes of dense clouds 
that are similar to stripping timescales, and also due to processes 
other than direct stripping, meaning some form of cloud destruction 
after the surrounding low density gas is stripped. Such processes may 
contain ablation, which includes Kelvin-Helmholtz instabilities, and 
self-destruction of the clouds by supersonic motion of the galaxy 
through the Virgo ICM. 

\item The surviving central molecular gas reservoir makes it possible 
for a nuclear star cluster to eventually form, even after most of the 
galaxy's gas has been stripped.

\item We have suggested that ram pressure may act as an external agent 
that, together with the thermal pressure of the ICM and assisted by 
radiative cooling, compresses the stripped dense clumps and thus may 
trigger star formation in the tail of the galaxy. 
\end{enumerate}

\acknowledgement
We acknowledge support by the projects RVO:67985815, M100031203 of the 
Academy of Sciences of the Czech Republic, GA~\v CR P209/11/P699, and 
GA~\v CR P209/12/1795. Pavel J\'achym was a Fulbright fellow at 
Yale University in 2009--2010 when this project was prepared. Ming Sun 
is supported by NASA grants GO2-13102X, GO0-11145C, and GO1-12103A. We 
thank H. Crowl and S. Tonnesen for useful comments. The scientific 
results reported in this article are based in part on observations made 
by the Chandra X-ray Observatory. This research has made use of the 
NASA/IPAC Extragalactic Database (NED) which is operated by the Jet 
Propulsion Laboratory, California Institute of Technology, under 
contract with the National Aeronautics and Space Administration, and of 
the HyperLeda database (http://leda.univ-lyon1.fr). 

\bibliographystyle{aa}
\bibliography{jachym-ic3418}
\end{document}